\documentclass[12pt,tightenlines,eqsecnum,floats,aps,amsmath,amssymb,nofootinbib,prd,showpacs]{revtex4}

\usepackage{setspace}
\usepackage{amsmath,amssymb,amsfonts,amsthm}
\usepackage{graphicx}
\usepackage{enumerate} 
\usepackage{colordvi} 

\def\be{\begin{equation}}
\def\ee{\end{equation}}
\def\ba{\begin{eqnarray}}
\def\ea{\end{eqnarray}}

\def\H{{\cal H}}
\def\Hkg{\H_{\rm kin}^{\rm grav}}
\def\Hk{\H_{\rm kin}^{\rm total}}
\def\Hkwdw{\H_{\rm kin}^{\rm wdw}}

\def\Hp{\H_{\rm phy}}

\def\Hpwdw{\H_{\rm phy}^{\rm wdw}}

\def\h{\hat }

\def\tr{{\rm Tr}}

\def\su{{\rm su}}
\def\SU{{\rm SU}}

\def\Tr{{\rm Tr\,}}

\def\cm{\rm cm}

\def\a{\alpha}
\def\b{\beta}
\def\g{\gamma}
\def\lp{{\ell}_{\rm Pl}}
\def\lo{{\ell}_o}
\def\R{\mathbb{R}}
\def\S{\mathbb{S}}
\def\Z{\mathbb{Z}}

\def\q{{}^o\!q}
\def\e{{}^o\!e}
\def\w{{}^o\!\omega}

\newcommand{\ket}[1]{\ensuremath{|#1\rangle}}
\newcommand{\ip}[2]{{\langle#1\,|\,#2\rangle}}
\newcommand{\rcr}{\rho_{\mathrm{crit}}}

\newcommand{\heff}{{\cal H}_{\mathrm{eff}}}

\newcommand{\p}{\partial}

\newcommand{\f}{\frac}

\newcommand{\ep}{\epsilon}

\def\lo{\ell_o}
\def\sfnsq{\sin^2\big(\f{\mb \lo}{2}\big)}
\def\sfnsqc{\sin^2\,\mb\big(c - \f{\lo}{2}\big)}
\def\sfn{\sin\,\mb\big(c - \f{\lo}{2}\big)}
\def\cfn{\cos\,\mb\big(c - \f{\lo}{2}\big)}

\def\chifn{\sfnsq - (1 + \gamma^2) \f{\mb^2 \lo^2}{4}}
\def\rmin{\rho_{\mathrm{min}}}
\def\rmax{\rho_{\mathrm{max}}}

\usepackage{enumerate}

\usepackage{colordvi}
\newcounter{mnotecount}[section]

\newcommand{\comment}[1]{}

\newcommand{\fs}[2]{{\textstyle\frac{#1}{#2}}} 
\def\vep{\varepsilon}

\def\f{\frac}

\def\La{\mathcal{L}}

\def\dd{\textrm{d}}
\def\ub{\underbar}
\def\ul{\underline}
\def\WDW{WDW\,\,}
\def\t{\tilde}

\def\mb{\bar \mu}

\def\bq{\begin{eqnarray}}
\def\eq{\end{eqnarray}}
\begin{document}
\preprint{\vbox{\baselineskip=12pt \rightline{IGPG-06/06-}
}}

\title{Loop quantum cosmology of $k$=1 FRW models}

\author{Abhay Ashtekar${}^{1,2}$}
\email{ashtekar@gravity.psu.edu}
\author{Tomasz Pawlowski${}^{1}$}
     \email{pawlowsk@gravity.psu.edu}
    \author{Parampreet Singh${}^{1}$}
    \email{singh@gravity.psu.edu}
    \author{Kevin Vandersloot${}^{1,3}$}
    \email{Kevin.Vandersloot@port.ac.uk}
    \affiliation{${}^{1}$Institute for Gravitational Physics and
Geometry, Physics Department, Penn State, University Park, PA
16802, U.S.A.\\
${}^2$ Institute for Theoretical Physics, University of Utrecht,
Princetonplein 5, 3584 CC Utrecht, The Netherlands\\
${}^3$ Institute of Cosmology and Gravitation, Mercantile
House,\\Portsmouth University, Portsmouth PO1 2EG, U.K.}

\begin{abstract}

The closed, $k$=1, FRW model coupled to a massless scalar
field is investigated in the framework of loop quantum cosmology
using analytical and numerical methods. As in the $k=0$ case, the
scalar field can be again used as emergent time to construct the
physical Hilbert space and introduce Dirac observables. The
resulting framework is then used to address a major challenge of
quantum cosmology: resolving the big-bang singularity while
retaining agreement with general relativity at large scales. It is
shown that the framework fulfills this task. In particular, for
states which are semi-classical at some late time, the big-bang is
replaced by a quantum bounce \emph{and} a recollapse occurs at the
value of the scale factor predicted by classical general
relativity. Thus, the `difficulties' pointed out by Green and
Unruh in the $k$=1 case do not arise in a more systematic
treatment. As in $k=0$ models, quantum dynamics is deterministic
across the deep Planck regime. However, because it also retains
the classical recollapse, in contrast to the $k=0$ case one is now
led to a cyclic model. Finally, we  clarify some issues raised
by Laguna's recent work addressed to computational physicists.

\end{abstract}


\maketitle

\section{Introduction}
\label{s1}

The spatially flat, isotropic, $k$=0 model was recently
investigated in detail in the setting of loop quantum cosmology
(LQC) \cite{aps1,aps2,aps3}.  That investigation introduced a
conceptual framework and analytical and numerical tools to
construct the \emph{physical sector} of the quantum theory. These
methods enabled one to systematically explore the effects of
quantum geometry both on the gravitational and matter sectors and
to extend previous results in LQC. The purpose of this paper is to
use the `improved dynamics' of \cite{aps3} to carry out a similar
analysis for the closed, $k$=1 model, coupled to a massless scalar
field. In our presentation, we will skip those constructions,
proofs and arguments which are direct analogs of the ones used in
the $k$=0 case in \cite{aps1,aps2,aps3}. The focus will be on the
differences from the existing treatments of the $k$=1 model and
also our earlier analysis of the $k$=0 model.

Although current observations favor spatially flat models, the
$k$=1 closed model is of considerable conceptual and technical
interest. On the conceptual side, an outstanding problem in
quantum cosmology ---and indeed in full quantum gravity--- is
whether one can construct a framework that cures the
short-distance difficulties faced by the classical theory near
singularities, while maintaining an agreement with it at large
scales. By their very construction, perturbative and effective
descriptions have no problem with the second requirement. However,
physically their implications can not be trusted at the Planck
scale and mathematically they generally fail to provide a
deterministic evolution across the putative singularity. In loop
quantum gravity (LQG) the situation is just the opposite. Quantum
geometry gives rise to new discrete structures at the Planck scale
that modify the classical theory in such a way that, at least in
simple models, space-like singularities of general relativity are
resolved. However, since the emphasis is on background
independence and non-perturbative methods, a priori it is not
clear whether the theory also has a rich semi-classical sector. Do
the novel dynamical corrections unleashed by the underlying
quantum geometry naturally fade away at macroscopic distances or
do they have unforeseen implications that prevent the theory from
reproducing general relativity at large scales? While there is
recent progress which indicates that LQG does admit a viable
semi-classical sector near Minkowski space-time \cite{cretal},
further evidence is needed in other contexts such as cosmological
settings.

The `improved dynamics' of \cite{aps3} successfully addressed this
issue in the $k$=0 case. However, because classical recollapse,
the $k$=1 model provides a more stringent tests. In particular
using numerical evolutions of equations used in the early LQC
works \cite{closed}, Green and Unruh \cite{gu} had concluded that
there may be a key difficulty: the LQC evolution appeared not to
reproduce the recollapse predicted by general relativity. Since
curvatures at the epoch of the classical recollapse are very
small, this feature appeared to indicate that LQC would deviate
from general relativity in perfectly ordinary situations. Thus,
although the LQC equations used in the Green-Unruh analysis did
cure the ultraviolet difficulties by resolving the singularity in
the sense of \cite{mb1}, they appeared not to have a viable
infrared behavior. Can this difficulty be resolved? Or, does the
situation in the $k$=1 model indicate that LQG may not admit a
good semi-classical sector in this cosmological setting? We will
employ methods developed in the $k$=0 case \cite{aps3} to probe
this issue in the \emph{physical sector} of the quantum theory. A
systematic analysis will show that not only is the big-bang
singularity resolved but the quantum evolution in fact faithfully
mirrors the predictions of general relativity, including the
re-collapse, when the curvature is small compared to the Planck
scale.

A second conceptual issue is whether the general features of the
Planck scale physics found in the $k$=0 case in \cite{aps3} are
robust. For example, there the Friedmann equation $(\dot{a}/a)^2 =
8\pi G\, \rho /3$ is replaced in LQC by the quantum corrected
equation $(\dot{a}/a)^2 = (8\pi G\, \rho /3)(1 -\rho/\rcr)$ where
$\rcr \approx 0.82 \rho_{\rm Pl}$. The correction comes with a
negative sign, making it possible for $\dot{a}$ to vanish
---triggering a bounce--- when the matter density reaches a
critical value, $\rcr$. One then has a pre-big-bang branch joined
on to the current post-big-bang branch by a deterministic
evolution, both behaving classically when the density is low
compared to $\rcr$. Does this feature survive in the $k$=1 case or
does space-time simply become fuzzy near and to the `past' of the
big-bang? Is there only one cycle in which the universe resembles
our own? Is the value of the critical density $\rho_{\rm max}$
at the bounce point essentially the same as in the $k$=0 model or
does it depend on the spatial topology? Our analysis will show
that the big-bang and the big-crunch are replaced by a quantum
bounce leading, in a precise sense, to a cyclic quantum universe.
Furthermore, the value of $\rho_{\rm max}$ is robust so long as
the maximum radius of the universe attains a macroscopic size.

On the technical side, this model also provides a number of
challenges. In LQG, the configuration variable is a connection
$A_a^i$, related to the spin connection $\Gamma_a^i$ (determined
by the spatial triad) and the extrinsic curvature $K_a^i$ via
$A_a^i = \Gamma_a^i + \gamma K_a^i$ where $\gamma$ is the
Barbero-Immirzi parameter. However, because of certain technical
difficulties with the methods used, so far the $k$=1 case has been
treated in the literature by regarding the extrinsic curvature
$K_a^i$ as a connection and constructing holonomies from it
\cite{closed}. Because of gauge fixing this is a mathematically
viable strategy. However as emphasized in, e.g. \cite{gu}, to make
closer contact with the full theory, it is desirable to construct
the theory using connections $A_a^i$. So an important question
arises: Can the improved dynamics of \cite{aps3} overcome the
difficulties faced in the earlier treatments, allowing us to
formulate the theory in terms of $A_a^i$? We will find that the
answer is in the affirmative. A second technical challenge arises
in the definition of the operator representing the field strength
$F_{ab}^i$ that features in the expression of the  Hamiltonian constraint.
In LQC, this operator is constructed from holonomies \cite{abl}.
In the $k$=0 model, one can use the integral curves of the (right
and left-invariant) fiducial triad $\e^a_i$ to build the holonomy
loops. In the $k$=1 case, $\e^a_i$ (is only left invariant and)
satisfies the commutation relations of ${\rm so(3)}$. Hence none
of the three pairs of triad vectors is surface forming. If one
uses a general loop which is not tangential to these triads,
holonomies (fail to be almost periodic functions of connections
and) do not lead to well-defined operators in the quantum theory.
Thus, finding suitable loops poses an interesting technical
challenge. We resolve this issue.%
\footnote{This issue was resolved independently by the Warsaw
group \cite{warsaw}. Their method of evaluating the holonomy is
more intrinsic, elegant and insightful.}
Finally, to numerically solve the quantum Hamiltonian constraint,
one has to find normalizable eigenfunctions of a certain
difference operator $\Theta$. Recall that in the case of, e.g.,  a
simple harmonic oscillator, the differential operator representing
the Hamiltonian admits an eigenfunction with any real number as
its eigenvalue. Normalizable eigenfunctions exist only for
discrete eigenvalues and thus constitute a set of zero measure in
the space of all eigenfunctions. In the $k$=1 case, the situation
is similar with the operator $\Theta$. However, now the
normalizable eigenfunctions and their eigenvalues are not known
analytically and, since they constitute a set of zero measure, one
has to exercise considerable care in singling them out
numerically.

Insights gained from the resolution of these conceptual and
technical issues are likely to be important as one considers more
and more general situations and develop new tools to tackle the
Hamiltonian constraint of full LQG.

The paper is organized as follows. In section \ref{s2} we
summarize the basics of the classical and quantum theories for our
model, arriving at the form of the Hamiltonian constraint
operator. In section \ref{s3} we sketch the Wheeler-DeWitt (WDW) theory and in
\ref{s4} we introduce the physical sector of the theory in LQC.
Numerical simulations are discussed in section \ref{s5}. In
section \ref{s6} we summarize effective equations of the
semi-classical sector, list the main numerical results and compare
them with predictions of the effective theory. We also clarify
some issues that have been raised by a numerical analysis by
Laguna \cite{pl} and remove confusion caused by an unfortunate use
of terminology there. Section \ref{s7} places these results in a
broader context and discusses their relation with that by Green
and Unruh \cite{gu}. Appendix \ref{a1} summarizes some background
material on the fiducial geometrical structures used in $k$=1
models with the so-called `Bianchi IX' symmetries.

\section{THE BASIC FRAMEWORK}
\label{s2}

This section is divided into two parts. In the first we summarize
the classical theory and in the second we present the basics of
the quantum theory.

\subsection{Classical Theory}
\label{s2.1}

Space-time manifolds under consideration are of the form $M\times
\R$, where $M$ has the topology of a 3-sphere, $\S^3$. As
explained in appendix \ref{a1}, one can identify $M$ with the
symmetry group $\SU(2)$ (which ensures spatial homogeneity and
isotropy) and endow it with a fixed fiducial basis of 1-forms
$\w_a^i$ and vectors $\e^a_i$. The resulting fiducial metric is
\be \q_{ab}:= \w_a^i\, \w_b^j\, k_{ij}, \quad\quad  \hbox{\rm
$k_{ij}$:\,\,\,  the Cartan-Killing metric on $\su(2)$.} \ee
$\q_{ab}$ turns out to be the metric of the round 3-sphere with
radius $a_o=2$ (rather than $a_o=1$). The volume of $M$ w.r.t. this
fiducial metric $\q_{ab}$ is $V_o = 2\pi^2a_o^3 = 16\pi^2$ and the
scalar curvature is ${}^o\!R = 6/a_o^2 = 3/2$. We shall set $\lo :=
V_o^{1/3}$. (For details, see Appendix \ref{a1}).

To facilitate comparison with the spatially flat, $k$=0 case and
to directly use results of \cite{aps1,aps2,aps3} obtained in that
case, we will set our conventions in a parallel fashion. The
dynamical, isotropic homogeneous connections and triads will be
parameterized by $c$ and $p$ respectively:
\be \label{ps} A_a^i = c\, \lo^{-1}\,\, \w_a^i, \quad\quad E^a_i =
p\, \lo^{-2}\,\sqrt{\q}\,\, \e^a_i\, . \ee
As in the $k$=0 case $c$ is dimensionless while $p$ has dimensions
of area and the density weight of $E^a_i$ is absorbed in the
determinant of the fiducial metric. At the point $(c,p)$ of the
phase space, the physical 3-metric $q_{ab}$ and the extrinsic
curvature $K_{ab}$ are given by:
\be q_{ab} = |p|\,\, \lo^{-2} \, \q_{ab}\quad {\rm and}\quad
\g\,K_{ab} = (c-\textstyle{\frac{\lo}{2}}) \,\,
|p|^{\fs{1}{2}}\,\, \lo^{-2} \, \q_{ab} \ee
The corresponding physical volume of $M$ is $|p|^{\fs{3}{2}}$. The
scale factor ${a}$ associated with a physical metric $q_{ab}$ is
generally expressed via $q_{ab} = {a}^2 \,
{}^o\!\underbar{q}_{ab}$ where ${}^o\! \underbar{q}_{ab}$ is the
\emph{unit} 3-sphere metric. Then, the scale factor is related to
$p$ via $|p| =  a^2 \lo^2 / 4$. As usual $p$ can take both
positive and negative values, the change in sign corresponds to a
flip in the orientation of the triads $\e^a_i$ which leaves the
physical metric $q_{ab}$ invariant.

Expressions of the total action, the symplectic structure, and the
Hamiltonian constraint require an integration over $M$. In the
$k$=0 case the underlying manifold is non-compact ($\R^3$), whence
the naive integrals would simply diverge. Therefore, irrespective
of which quantization scheme one uses, one has to fix a fiducial
cell with finite volume, say ${\cal V}_o$ (w.r.t. the fiducial
flat metric), and restrict all integrations to this cell. In the
$k$=1 case, $M$ is compact and the introduction of a cell is
unnecessary. Nonetheless, in many of the key equations our $V_o$
plays the role of the volume ${\cal V}_o$ of the  cell in the
$k$=0 case. In both cases, the fundamental Poisson bracket is:
\be \label{pb} \{c,\, p\} = \f{8\pi G\g}{3}   \ee
where $\g$ is the Barbero-Immirzi parameter. Finally, using the
fact that the Cartan orthonormal triad $\omega_a^i$ on a 3-sphere
\emph{of radius $a_o$} satisfies the identity
\be \label{cartan} \dd \omega^k + \frac{1}{a_o}\, \ep_{ij}{}^k
\omega^i\wedge \omega^j =0\, , \ee
it is straightforward to calculate the field strength $F_{ab}^k$
of the connection $A_a^i$ on $M$
\be \label{F1} F_{ab}^k = \lo^{-2}\, \left[ c^2 - c\,\lo
({\f{2}{a_o}})\right] \epsilon_{ij}{}^k\, \w_a^j\,
\w_b^k\, . \ee
Our fiducial co-triad $\w_a^i$ corresponds to $a_o=2$. But we have
refrained from using this numerical value in (\ref{F1}) to clarify
the relation with the $k$=0 case. To pass to this case, one has to
set $V_o$ equal to the volume ${\cal V}_o$ of the fiducial cell
there but take the limit as $a_o$ goes to infinity. In this limit,
the fiducial co-triad $\w_a^i$ goes to the (flat) fiducial
co-triad used in the $k$=0 case (see (\ref{cartan})) and the field
strength $F_{ab}^k$ goes over to its value ${\cal V}_o^{-2/3}
c^2\, \epsilon_{ij}{}^k\, \w_a^i \,\w_b^j$ in the $k$=0 case
\cite{abl,aps1,aps2,aps3}.

As in the $k$=0 models, we have completely gauge fixed the Gauss
and the diffeomorphism constraints by fixing $\w_a^i$ and $\e^a_i$
and using the parametrization (\ref{ps}) of the phase space
variables $A_a^i, E^a_i$. So, we are left only with the
Hamiltonian constraint. The gravitational part of this constraint
is given by \cite{alrev,ttbook}:
\be \label{ham1} C_{\mathrm{grav}} = \int_{M} \dd^3 x\,
\left[\ep^{ij}{}_{k} \, e^{-1}\, E^{ai}E^{bj}\, \, F_{ab}^k \,-\,
2(1+\g^2)\,\, e^{-1}\,E^{ai}E^{bj}\, \, K_{[a}^i K_{b]}^j\right]\,
, \ee
where $e := \sqrt{|\mathrm{det}E|}$ and $K_a^i = K_a{}^b\,\w_b^i$
is the extrinsic curvature, and where, as is usual in
mini-superspace analysis, we have set the lapse equal to $1$.
Using the fact that $A_a^i$ is related to the spin-connection
$\Gamma_a^i$ (of the physical triad $e^a_i$) and the extrinsic
curvature $K_a^i$ through $A_a^i = \Gamma_a^i + \g K_a^i$, it is
convenient to express the second term in the integrand of
(\ref{ham1}) in terms of the curvature $F_{ab}^k$ of $A_a^i$ and
the curvature $\Omega_{ab}^k$ of $\Gamma_a^i$:
\be E^{ai}E^{bj}\, \, K_{[a}^i K_{b]}^j = \f{1}{2\g^2}\,
\ep^{ij}{}_k \, E^{ai}E^{bj}\,\left( F_{ab}^k -
\Omega_{ab}^k\right)\, .\ee
In the $k$=0 case, the spatial curvature $\Omega_{ab}^k$ vanishes
and the extrinsic curvature term in (\ref{ham1}) reduces to a
multiple of the first term involving $F_{ab}^k$. In full general
relativity, by contrast, while $\Omega_{ab}^i$ is determined by
the momenta $E^a_i$, its expression is rather complicated and this
strategy of handling the extrinsic curvature terms, by itself,
does not simplify matters. The situation in general homogeneous
models \cite{closed} as well as black hole interiors 
\cite{ab1} is in between the two: Although $\Omega_{ab}^k$ is
non-zero, its expression is simple and manageable. In the $k$=1
model now under consideration one has:
\be \Omega_{ab}^k = - \f{1}{4}\, \epsilon_{ij}{}^k \,\,\w_a^i\,
\w_b^j\, . \ee
Therefore, the gravitational part of the constraint reduces to:
\ba C_{\mathrm{grav}} &=& -\f{1}{\g^2}\,\, \int_{M} \dd^3 x\,
\ep^{ij}{}_{k} \, e^{-1}\, E^{a}_{i}E^{b}_{j}\, \, \bigg[F_{ab}^k
\,-\, \big(\f{1+\g^2}{4}\big)\, {}^o\!\epsilon_{ab}{}^c\,\,
\w_c^k\bigg]
\label{ham2}\\
&=& - \f{6\sqrt{p}}{\g^2}\, \bigg[ \big(c - \f{\lo}{2}\,
\f{2}{a_o}\big)^2 + \f{\g^2\lo^2}{4}\, \f{4}{a_o^2}
\label{ham3}\bigg]\ea
where, in the last step we have used the expression (\ref{F1}) of
$F_{ab}$. In the $k$=1 case now under consideration, $a_o=2$.
However, as indicated above, we did not substitute this numerical
value because results for the $k$=0 case \cite{aps1,aps2,aps3} can
be recovered by setting $a_o=\infty$ (and $\lo = {\cal V}_o^{1/3}$).
However, since $a_o$ always occurs in the combination $\lo/a_o$ in the
Hamiltonian constraint, we can set $a_o=2$ throughout and recover
the $k$=0 results simply by setting $\lo=0$. In what follows, we
will adopt this strategy.

\textbf{Remark:} In the above construction, we began with a
fiducial triad $\e^a_i$ and a co-triad $\w_a^i$ adapted to a
3-sphere of radius $a_o$=2. Therefore, our construction may appear
to be tied to that choice. Had we used a 3-sphere of radius $a_o=
2\lambda$, the fiducial triad and the co-triad we have rescaled
via $\e^a_i \rightarrow \lambda^{-1} \w_a^i$ and $\w_a^
\rightarrow \lambda \w_a^i$. It is easy to check that the
variables $c,p$ parameterizing the physical fields $(A_a^i,
E^a_i)$ are left unchanged. Hence the entire framework is
invariant under this rescaling `gauge' freedom.

\subsection{Quantum kinematics and the Hamiltonian constraint}
\label{s2.2}

To pass to the quantum theory, following Dirac one first
constructs a kinematical description. As in the $k$=0 case
\cite{abl,aps1,aps2,aps3} the kinematical Hilbert space $\Hkg$ is
the space $L^2(\R_{\rm Bohr}, d\mu_{\rm Bohr})$ of square
integrable functions on the Bohr compactification of the real
line. To specify states concretely, it is convenient to work with
the representation in which the operator $\hat{p}$ is diagonal.
Eigenstates of $\hat{p}$ are labeled by a real number $\mu$ and
satisfy the orthonormality relation:
 \be \ip{\mu_1}{\mu_2} = \delta_{{\mu_1},\, {\mu_2}}\, . \ee
Since the right side is the Kronecker delta rather than the Dirac
delta distribution, a typical state in $\Hkg$ can be expressed as
a countable sum; $\ket\Psi = \sum_n c^{(n)} \ket{\mu_n}$ where
$c^{(n)}$ are complex coefficients and the inner product is given
by
\be \langle \Psi_1| \Psi_2 \rangle = \sum_n \, \bar c_1^{(n)} \,
c_2^{(n)} ~. \ee
The fundamental operators are $\hat{p}$ and $\widehat{\exp
i\lambda(c/2)}$:
\be \hat{p}\,\ket{\mu} = \f{8\pi \g \lp^2}{6}\, \mu\, \ket{\mu}
\quad {\rm and} \quad \widehat{\exp {\f{i\lambda
c}{2}}}\,\ket{\mu} = \ket{\mu+\lambda}\ee
where $\lambda$ is any real number. From the discussion of the
classical theory of section \ref{s2.1} it follows that the
physical volume operator of $M$ is given by: $\h{V} \,=\,
|\hat{p}|^{{3}/{2}}$.

Of special interest to us are holonomies of the connection $A_a^i$
along the integral curves of our fiducial triads $\e^a_i$. The
holonomy $h_k^{(\lambda)}$ along the segment of (directed) length
$\lambda \lo$, tangential to $e^a_k$ is given by:%
\footnote{Here $\mathbb{I}$ is the unit $2\times 2$ matrix and
$\tau_k$ is a basis in the Lie algebra $\su(2)$ satisfying
$\tau_i\, \tau_j = \f{1}{2}\ep_{ijk}\tau^k - \f{1}{4}
\delta_{ij}$. Thus, $2i \tau_k = \sigma_k$, where $\sigma_i$ are
the Pauli matrices. The directed length is positive if the line
segment is oriented along $\e^a_k$ and negative if it is oriented
in the opposite direction.}
\be \label{hol} {h^{(\lambda)}_k} = \cos \f{\lambda
c}{2}\,\mathbb{I} + 2 \sin \f{\lambda c}{2}\, \tau_k \, .\ee
The corresponding holonomy operator has the action:
\be \h{h}^{(\lambda)}_k\,\ket{\mu} = \f{1}{2}\, \left(
\ket{\mu+\lambda} + \ket{\mu -\lambda} \right) \mathbb{I} +
\f{1}{i}\, \left( \ket{\mu+\lambda} - \ket{\mu -\lambda} \right)
\tau_k  ~.\ee
However, just as there is no operator corresponding to the
connection itself in full LQG \cite{alrev,ttbook}, there is no
operator $\hat{c}$ on $\Hkg$ \cite{abl}.

To describe quantum dynamics, we have to first introduce a
well-defined operator on $\Hkg$ representing the Hamiltonian
constraint $C_{\rm grav}$. Since there is no operator
corresponding to $c$ itself, as in the $k$=0 case we will use the
integral expression (\ref{ham2}) of the constraint. For the passage to
quantum theory, one has to first express this classical constraint
in terms of the elementary variables $p$ and $h_k^{(\lambda)}$ and
then replace them with operators $\h{p}$ and
$\h{h}_k^{(\lambda)}$. As in the full theory \cite{tt,ttbook}, the
term involving triads becomes the following operator
\cite{abl,aps2,aps3}
\be \label{cotriad} \ep_{ijk}\, \widehat{e^{-1}\,E^{aj}E^{bk}}\,
=\, \sum_k \f{({\rm sgn}\,p)}{2i\hbar\,\pi\gamma G\lambda\, \lo}\,
\, {}^o\!\ep^{abc}\,\, \w^k_c\,\, \Tr\left( \h{h}_k^{(\lambda)}\,
[\h{h}_k^{(\lambda)}{}^{-1}, \h{V}]\, \tau_i\right) \ee
where $\h{V}= |\h{p}|^{3/2}$ is the volume operator.

To define $\h{F}_{ab}{}^k$ in (\ref{ham2}), as in
\cite{abl,aps1,aps2,aps3} we use the standard relation between the
holonomies and field strengths. Because of homogeneity, the
components $\e^a_i\,\e^b_j\, F_{ab}{}^k$ are constant on $M$. They
can be evaluated by considering a square loop $\Box_{ij}$ starting
and ending at any point $x$, with tangent vectors $\e^a_i$ and
$\e^b_j$ at $x$, and then taking the limit as the area enclosed by
the loop shrinks to zero. In the quantum theory, since there is no
operator corresponding to the connection $c$, the limiting
operator does not exist. As discussed in detail in
\cite{abl,aps1,aps2,aps3}, this is a manifestation of the quantum
nature of geometry, i.e., a reflection of the fact that the area
operator has purely discrete spectrum. As in \cite{aps3} our
strategy is to shrink the loop only until its \emph{physical} area
equals the `area gap' $\Delta$, i.e., the minimal non-zero
eigenvalue of the area operator.

Now, in the $k$=0 case, the edges of $\Box_{ij}$ can be taken to
be the integral curves of the triad vector fields $\e^a_i$ and
$\e^b_j$. In the $k$=1 model, however, the $\e^a_i$ satisfy the
commutation relations of $\su(2)$; they  do not commute. Therefore
their integral curves can not provide the desired closed loop
$\Box_{ij}$. In the existing literature, a closed loop is formed
by simply adding a fifth edge \cite{closed}. However, this
strategy is not viable: Since the five edges do not span an
unambiguous 2-surface, the notion of the area enclosed by the loop
has no obvious meaning. However, note that while $\e^a_i$ are the
`left-invariant' vector fields, $M$ also admits three `right
invariant' vector fields $\xi^a_i$ (see Appendix \ref{a1}). These
are the symmetry fields: they also satisfy the commutation
relations of $\su(2)$, act simply and transitively on $M$ and
their action leaves each of our fiducial triads $\e^a_i$ and
co-triads $\w_a^i$ invariant. Since they commute with $\e^a_i$,
one can form the desired closed loops $\Box_{ij}$ by first moving
from $x$ along the integral curve of say $\e_i$ then $\xi_k$ then
along $-\e_i$ and then along $-\xi_k$ (where $\xi_k$ is chosen to
coincide with $\e_j$ at $x$). An explicit realization  of this
procedure is presented in Appendix \ref{a1} and a more geometric
construction appears in \cite{warsaw}. The final field strength
operator does not depend on whether the first segment is chosen to
be left invariant or right invariant and is
given by%
\footnote{Here and in what follows, in light of results of
\cite{ap,kv}, we have used the fundamental, $j=1/2$
representation. For a more detailed discussion, see
\cite{aps2,aps3}.}:
\ba \label{F2} \h{F}_{ab}^k &=& \lim_{Ar\, \Box_{ij}\rightarrow
\Delta}\,\,\, \f{1}{\lambda^2\lo^2} \,\,\left(\sin^2 \lambda(c
-{\lo}/{2}) - \sin^2 (\lambda\lo/2)\,\right)\,\, \ep_{ij}{}^k \,\,
\w_a^i\w_b^j\nonumber\\
&=& \f{1}{\mb\lo^2} \,\,\left(\sin^2 \mb(c -{\lo}/{2}) - \sin^2
({\mb\lo}/{2})\,\right) \,\, \ep_{ij}{}^k \,\, \w_a^i\w_b^j\, .
\ea
Here, as discussed in detail in \cite{aps3}, $\mb$ is a specific
function of $p$:
\be \label{mubar} \mb^2\,\, |p|\, =\, \Delta\, \equiv \,
(2\sqrt{3}\pi\g) \, \lp^2\, , \ee
and for notational simplicity we have dropped hats on operators
which are trigonometric functions of $c$. The fact that $\mb$ is a
function of $p$ rather than a constant arises from the requirement
that the \emph{physical} area of $\Box_{ij}$ be set equal to
$\Delta$. As explained in \cite{aps3}, this strategy mimics the
full theory in a well-defined sense and the resulting `improved
dynamics' is free of the drawbacks of older Hamiltonian constraint
of LQC. As in \cite{abl,aps1,aps2,aps3}, the viewpoint is that at
the fundamental level ---i.e. at the Planck scale--- the field
strength operator is non-local and the usual local classical
expression arises only on coarse graining in semi-classical
states.

To obtain the explicit action of $\h{F}_{ab}{}^k$ on $\Hkg$, one
has to face two complications. The first arises already in the
$k$=0 case. The operator $\h{F}_{ab}{}^k$ for $k$=0 can be
recovered by setting $\lo=0$. It depends on the connection only
through $\sin(\mb c)$. Since $\mb$ is itself a function of
$\h{p}$, the action of $\sin \mb c$ on $\Hkg$ is rather subtle. As
discussed in detail in \cite{aps3}, it is simplest to express it
by going to a basis $\ket{v}$ which is better adapted to the
volume operator $\h{V}$:
\be \label{V} \h{V}\ket{v} =  (\f{8\pi\g}{6})^{\f{3}{2}}\,\,
\f{|v|}{K}\,\,\lp^3 \,\, \ket{v} \ee
where the dimensionless label $v$ ---the eigenvalue of $\h{V}$
apart from a constant--- is related to the dimensionless label
$\mu$ ---the eigenvalue of $\h{p}$ apart from an overall
constant--- via
\be \label{v} v = K \,{\rm sgn}(\mu)\, |\mu|^{\f{3}{2}}, \quad
{\rm where} \quad K =  \f{2\sqrt{2}}{3\sqrt{3\sqrt{3}}}\, . \ee
In this basis, ${\exp{i\mb c}}$ are simply the translation
operators:
\be \label{mbop2} {e^{i \f{\mb c}{2}}}\, \Psi(v) = \Psi(v+1), \ee
so that
\be \label{sine} {\sin({\mb c})}\, \Psi(v) = \f{1}{2i}\,
(\Psi(v+2) - \Psi(v-2)\,) .\ee

\medskip In the $k$=1 case we have an added complication:
(\ref{F2}) contains $\sin \mb(c-\lo/2)$ rather than $\sin (\mb
c)$. This difference can be handled as follows: On
$\Hkg$
\be \label{shift}{\sin (\mb c - \f{\lo}{2})}\, \Psi(v) \,=\,
e^{i\lo f}\, \sin\mb c \,\, e^{-i\lo f} \, \Psi(v), \quad {\rm where}\quad f =
\f{{\rm sgn} \, v}{4}\,\, \big|\f{v}{K}\big|^\f{2}{3}\, . \ee
Note that since $f(v)$ is continuous everywhere (including the
point $v=0$), the operator ${\exp i\lo f}$ is well-defined and
unitary on all of $\Hkg$.

We now have operators corresponding to each term in the integrand
of the gravitational part (\ref{ham2}) of the Hamiltonian
constraint. Using (\ref{cotriad}), (\ref{F2}), (\ref{sine}) and
(\ref{shift}) in (\ref{ham2}) we obtain:
\ba \label{ham4} \h{C}_{\mathrm{grav}}\, \Psi(v) \,= &&e^{if\lo}
\,\,\sin\mb c\, \h{A} \sin\mb c\,\, e^{-i\lo f}\,\, \Psi(v)\nonumber\\
&-&\big[\sin^2\f{\mb\lo}{2} - \f{\mb^2\lo^2}{4} - \f{\lo^2}{9|K^2
v|^{\fs{2}{3}}}\big]\,\, \h{A}\, \Psi(v) \ea
where, as in the $k$=0 analysis \cite{aps3} we have set
\be \label{A} \hat A \,\Psi(v)\, =\, - \f{27K}{4}
\,\sqrt{\f{8\pi}{6}} \, \f{\lp}{\gamma^{3/2}} \,|v| \,\,\, \big|
|v - 1| - |v + 1| \big|\, \,\, \Psi(v) ~. \ee
To summarize, there are two main subtleties in the passage from
spatially flat, $k$=0 models to the closed, $k$=1 ones. First, the
loop around which holonomy is computed has to be constructed using
both right and left invariant vector fields. Second, now the
connection dependence is in the operator $\sin \mb(c-\lo/2)$
rather than $\sin \mb c$ and one has to define its action on
$\Hkg$ using the unitary operators $\exp i\lo f$. When this is
done, the gravitational part of the Hamiltonian constraint is
symmetric and positive, given by (\ref{ham4}).

Finally, to write the complete constraint operator we also need
the matter part of the constraint. For the massless scalar field,
in the classical theory it is given by:
\be C_{\rm matt} = 8\pi G\,\, |p|^{-\f{3}{2}}\, p_\phi^2 \ee
As usual, the non-trivial part in the passage to quantum theory is
the function $|p|^{-3/2}$. However, as with the co-triad operator
(\ref{cotriad}), we can use the method introduced by Thiemann in
the full theory \cite{tt,ttbook}. This issue is discussed in
detail in \cite{aps3}. The final result is:
\be \label{inversevol}  \widehat{{|p|^{-\f{3}{2}}}} \Psi(v) =
\left(\f{6}{8 \pi \gamma \lp^2}\right)^{3/2}\, B(v)\, \Psi(v) \ee
where
\be \label{B} B(v) = \left(\f{3}{2}\right)^3 \, K\,\, |v| \,
\bigg| |v + 1|^{1/3} - |v - 1|^{1/3} \bigg|^3\, . \ee
It is self-adjoint on $\Hkg$ and diagonal in the eigenstates of
the volume operator.

Collecting these results we can express the total constraint
\be \label{hc1} \h{C}\, \Psi(v) = \left(\hat{C}_{\rm grav} +
\hat{C}_{\rm matt}\right)\, \Psi(v) = 0\, ,\ee
as follows:
\ba \label{hc2} \p^2_\phi \Psi(v,\phi) = &-& \Theta\, \Psi(v,\phi)
\nonumber\\
= &-& \Theta_o \Psi (v,\phi) + \f{\pi G}{2} [B(v)]^{-1} \Big[3K
(\sin^2(\f{\mb\lo}{2}) - \f{\mb^2\lo^2}{4})\, |v|\,
\nonumber\\
&-& \, \f{1}{3}\, \lo^2\g^2\,\,
\big|\f{v}{K}\big|^{\fs{1}{3}}\Big] \,\,\Big[ \big|\,
 |v-1|- |v+1|\,\big|\Big]\, \Psi(v,\phi)\, .\ea
Here, $\Theta_o$ is the operator that appears in the $k$=0 quantum
constraint in place of $\Theta$ \cite{aps3}:
\be \label{hc3} \Theta_o \Psi(v,\phi) =  - [B(v)]^{-1} \,
\left(C^+(v)\, \Psi(v+4,\phi) + C^o(v) \, \Psi(v,\phi) +C^-(v)\,
\Psi(v-4,\phi)\right) \ee
where the coefficients $C^\pm(v)$ and $C^o(V)$ are given by:
\ba \label{C} C^+(v) &=& \nonumber \f{3\pi K G}{8} \, |v + 2|
\,\,\,
\big| |v + 1| - |v +3|  \big|  \\
C^-(v) &=& \nonumber C^+(v - 4) \\
C^o(v) &=& - C^+(v) - C^-(v) ~. \ea
Thus, the $k$=1 quantum constraint has the same form as in the
$k$=0 case. As one would expect from the classical expression
(\ref{ham2}), the difference $\Theta-\Theta_o$ is diagonal in the
$v$-representation and vanishes when we set $\lo=0$.

In the remainder of the paper, we will work with the Hamiltonian
constraint (\ref{hc2}). As in the $k$=0 case \cite{aps2,aps3}, the
form of this constraint is similar to that of a massless
Klein-Gordon field in a static space-time, but now with an
additional static potential. $\phi$ is the analog of the static
time coordinate and the difference operator $\Theta$, of the
spatial Laplace-type operator plus the static potential. Hence,
the scalar field $\phi$ can again be used as `emergent time' in
the quantum theory. We will examine the operator $\Theta$ in some
detail in sections \ref{s4} and \ref{s5}. Finally, in the above
construction we made a factor ordering choice the motivations
behind which are the same as those discussed in \cite{aps3}. This
choice will facilitate comparison between the LQC results in the
$k$=1 and $k$=0 cases and yield the \WDW equation with its
`natural' factor ordering in the `continuum limit'.

\textbf{Remark:} Velhinho \cite{v} has pointed out that in quantum
kinematics it would suffice to consider the algebra generated by
$p$ and just two almost periodic functions of the connection,
$e^{i\mu_1 c}$ and $e^{i\mu_2 c}$, where $\mu_1/\mu_2$ is
irrational, because these functions already separate points of the
phase space. In the older, `$\mu_o$-evolution' \cite{abl,aps2},
this strategy would be natural; one could set $\mu_1 =1$ and
$\mu_2 =\mu_o \equiv 3\sqrt{3}/2$. However, as discussed in
\cite{aps2}, this evolution is not viable physically. The
`improved' $\mb$-evolution used in \cite{aps3} and in this paper
is free of those drawbacks. But since $\mb$ is now not a constant
but a function (\ref{mubar}) of $\mu$, the Velhinho kinematics
will not support the `improved quantum dynamics.'

\section{WHEELER DEWITT THEORY}
\label{s3}

In this section we will briefly discuss the \WDW limit of LQC in
which effects specific to quantum geometry in the difference
equation (\ref{hc2}) are ignored by letting the area gap go to
zero. This discussion will bring out the key role played by
quantum geometry near the big bang and the big crunch
singularities. The \WDW theory has its roots in geometrodynamics
which is insensitive to the choice of the triad orientation
---i.e., to the sign of $v$. Therefore, as in \cite{aps2,aps3}, we
will restrict ourselves to wave functions $\ul\Psi(v)$ which are
symmetric under the orientation reversal operator $\Pi$,
\be \label{pi} \Pi \,\,\ul\Psi(v,\phi) = \ul\Psi(-v,\phi)\, .\ee
where (and in what follows) we have denoted the \WDW analogs of
the LQC quantities with an underbar. As in \cite{aps3} we will
find that the scalar field $\phi$ can again be used as emergent
time and, in the resulting physical sector of the theory, the big
bang and the big crunch singularities persist in the \WDW limit.
Details of motivation and background material as well as underlying
assumptions and
technical steps involved in the \WDW limit can be found in
\cite{aps3}.

\subsection{The \WDW constraint and its general solution}
\label{s3.1}

To pass to the \WDW theory, one sends the area gap  to zero
and restricts oneself to \emph{smooth} wave functions
$\ul\Psi(v,\phi)$. As explained in \cite{aps2,aps3}, this
corresponds to taking the continuum limit of the difference
equation (\ref{hc2}).
The non-trivial part of the task lies in the limit of $\Theta_o$ since
the remainder is diagonal in $v$. This has already been completed
in \cite{aps3}. The required \WDW limit is then given by%
\footnote{The appearance of the Barbero-Immirzi parameter $\g$ in
the \WDW limit is an artifact of our conventions, i.e., definition
of $v$. If instead we use the eigenvalue $\bar{v}$ of the volume
operator, $\h{V} \ket{\bar{v}} = \bar{v}\lp^3\,\ket{\bar{v}}$, so
that $\bar{v} = (8\pi \g/6)^{\fs{3}{2}}\, (v/K)$, the \WDW
equation (\ref{wdw1}) would be manifestly independent of $\g$:
$\p_\phi^2 \ul\Psi(\bar{v},\phi) = 12\pi G (\bar{v}\,
\p_{\bar{v}})^2\, \ul\Psi (\bar{v}, \phi) - (G\lo^2/16\pi)
\bar{v}^{{4}/{3}}\,  \ul\Psi(\bar{v},\phi)$.}
\ba \label{wdw1}\p_\phi^2 \ul\Psi(v,\phi) &=& -\ul{\Theta}
\ul\Psi(v,\phi) \nonumber\\ &=& \, -{\ul{\Theta}}_o
\ul\Psi(v,\phi) - \frac{\pi G\lo^2\gamma^2}{3K^{\frac{4}{3}}}|v|^{\frac{4}{3}}\, \ul\Psi(v,\phi) \ , \ea
where, as in \cite{aps3},
\be {\ul{\Theta}}_o\ul\Psi(v,\phi):=-12\pi G\, (v\partial_v)^2\,
\Psi(v,\phi)\, .\ee
As explained in \cite{aps3}, this factor ordering is `covariant'
from the geometrodynamical perspective and coincides with the one
used in the older \WDW literature (see e.g. \cite{ck}).

The operator $\ul\Theta$ is self-adjoint and positive definite on
the Hilbert space $L^2(\R, \ub{B}(v)\dd v) \equiv L^2(\R,
(K/|v|)\dd v)$, where, as in \cite{aps3}, to facilitate comparison
with LQC results we have denoted the \WDW limit $K/|v|$ of $B(v)$
by $\ub{B}(v)$. The general solution of (\ref{wdw1}) can be
readily expressed in terms of the spectral family of $\ul\Theta$.
Let us begin by considering all eigenfunctions of the differential
operator $\ul\Theta$:
\be  -12\pi G\, (v\partial_v)^2\, \psi_{\omega}(v)
  + \frac{\pi G\lo^2\gamma^2}{3K^{\frac{4}{3}}}|v|^{\frac{4}{3}}
  \psi_{\omega}(v) \,= \, \omega^2 \psi_{\omega}(v)\, .
\ee
In their most general form, they can be expressed as linear
combinations of modified Bessel functions $\mathcal{I,K}$
\cite{ck}
\be
  \psi_{\omega}(v)\ =\ \alpha\,\, \mathcal{K}_{ik}\left(
    \frac{\lo\gamma}{4K^{\frac{2}{3}}}|v|^{\frac{2}{3}}
  \right)
  \ +\ \beta\,\, \mathcal{I}_{ik}\left(
    \frac{\lo\gamma}{4K^{\frac{2}{3}}}|v|^{\frac{2}{3}}
  \right) \ ,
\ee
where $k:= (3/16\pi G)^{\fs{1}{2}} \,\omega$\,\, and
$\alpha,\beta$ are constants. (We have used calligraphic letters
to denote the modified Bessel functions to avoid confusion with
the constant $K$ of (\ref{v})). As $x\rightarrow \infty$, the
function $\mathcal{I}_{ik}(x)$ grows exponentially whereas
$\mathcal{K}_{ik}(x)$ decays exponentially.

These properties of eigenfunctions imply that the spectrum of the
operator $\ul\Theta$ is continuous and that $\mathcal{I}_{ik}(x)$
cannot feature in its spectral decomposition. Therefore to obtain
this decomposition, we will set $\beta(k) =0$ and choose constants
$\alpha(k)$ such that the resulting eigenfunctions $\ub{e}_k(v) :=
\alpha(k) \mathcal{K}_{ik}(v)$ are orthonormal,
\be  \ip{\ub{e}_k}{\ub{e}_{k^\prime}} = \delta(k,k^\prime)\, . \ee
The Dirac distribution appears on the right hand side because the
spectrum of $\Theta$ is continuous. This may appear surprising
because $\ub{e}_{k}(v)$ decay for large $v$. However,
$\ub{e}_k(v)$ fail to have finite norm in $L^2(\R, (K/|v|) \dd v)$
because they have the following oscillatory form for small $v$
\be \mathcal{K}_{ik}\left(
\frac{\lo\gamma}{4K^{\frac{2}{3}}}|v|^{\frac{2}{3}} \right) \
\xrightarrow{|v|\ll k}\
a(\omega)\cos\left(\frac{\omega}{\sqrt{12\pi G}}\ln|v| +
\sigma(\omega) \right) \ , \ee
where $\sigma(\omega)$ is a constant (but $\omega$-dependent)
phase shift.

In the spatially open, $k$=0 models \cite{aps3}, the spectrum of
$\ul\Theta$ is 2-fold degenerate, reflecting the fact that in the
classical theory there are two sets of distinct solutions, one
perpetually expanding and the other perpetually contracting. In
the present case, this degeneracy is broken because
$\mathcal{I}_{ik}(v)$ diverge at large $v$. Its classical
counterpart is the fact that, because each solution has both
contracting and expanding epochs, we no longer have two distinct
sets of universes.

Eigenfunctions $\ub{e}_k(v)$ provide the standard spectral
decomposition on $L^2(\R, \ub{B}(v) \dd v)$:
\be \ul\Psi(v) = \int_0^\infty \dd k \t{{\Psi}}(k)\, \ub{e}_k(v)\,
. \ee
Hence, the general solution to (\ref{wdw1}) with smooth initial
data consisting of rapidly decreasing functions can be written as
\be \label{sol1} \ul\Psi(v, \phi) = \int_{0}^{\infty}\, \dd k \,
\t{{\Psi}}_{+}(k)\, \ub{e}_k(v) e^{i\omega \phi}+
\t{{\Psi}}_{-}(k)\, \bar{\ub{e}}_k(v) e^{-i\omega \phi}\,  \ee
for some suitably regular function $\t{{\Psi}}_\pm(k)$. Following
the terminology generally used in the Klein-Gordon theory, the
solution will be said to be of positive (resp. negative) frequency
if $\t{{\Psi}}_- (k)$ (resp. $\t{{\Psi}}_+ (k)$) vanishes.

As usual, the positive and negative frequency solutions satisfy
first order `evolution' equations, obtained by taking a
square-root of the constraint (\ref{wdw1}):
\be \label{freq1}\mp i\, \p_\phi \ul\Psi(v,\phi) =
\sqrt{\ul\Theta}\, \ul\Psi(v,\phi)\, . \ee
If $f(v)$ is the initial data for these equations at `time
$\phi=\phi_o$', the solutions are given by:
\be \label{sol2} \ul\Psi_\pm(v,\phi) = e^{\pm i \sqrt{\ul\Theta}\,
\,(\phi-\phi_o)}\,\, f(v)\, . \ee

\subsection{Physical sector of the \WDW theory}
\label{s3.2}

Solutions (\ref{sol1}) to the \WDW equation are not normalizable
in $\Hkwdw$ (because zero is in the continuous part of the
spectrum of the \WDW operator). Therefore, one has to use one of
the standard methods \cite{at,dm} to construct the physical
Hilbert space $\Hpwdw$. Since the procedure is completely
analogous to that used in \cite{aps2,aps3}, we will simply
summarize the final results.

$\Hpwdw$ consists of positive frequency solutions
$\ul\Psi(v,\phi)$ to (\ref{wdw1}) which are symmetric under the
reversal of the orientation of the triad, i.e. satisfy
$\ul\Psi(v,\phi) = \ul\Psi(-v, \phi)$, and which have a finite
norm w.r.t. the inner product:
\be \label{ip1} \langle\ul\Psi_1\,|\ul\Psi_2\, \rangle_{\rm phy} =
\int_{\phi=\phi_o}\!\! \dd v\, \ub{B}(v)\,
\bar{\ul\Psi}_1(v,\phi)\, \ul\Psi_2(v,\phi)\,  \ee
where for notational simplicity here (and in what follows) we have
dropped the subscript $+$ denoting positive frequency. On this
space, a useful complete set of Dirac observables is provided by
the momentum $\h{p}_\phi$ of the scalar field,
\be \label{dirac1}\hat{p}_\phi\, \ul\Psi(v, \phi) := -i\hbar
\f{\partial \ul\Psi(v,\phi)}{\partial \phi} \ee
and the operator $\h{|v|}_{\phi_o}$ corresponding to volume at the
emergent time $\phi=\phi_o$,
\be \label{dirac2}{|\h{v}|_{\phi_o}}\,\, \ul\Psi(v,\phi) =
e^{i\sqrt{\ul{\Theta}}\,\,(\phi-\phi_o)}\, |v|
\,\ul\Psi(v,\phi_o). \ee

\begin{figure}[]
  \begin{center}
$a)$\hspace{8cm}$b)$
    \includegraphics[width=3.2in,angle=0]{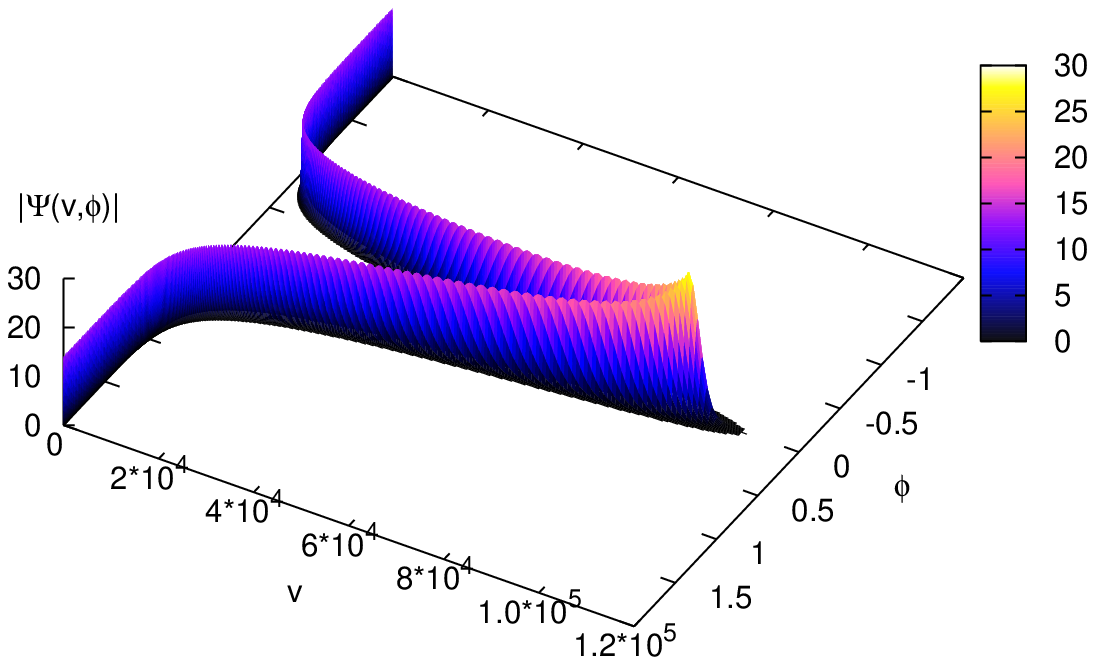}
\includegraphics[width=3.2in,angle=0]{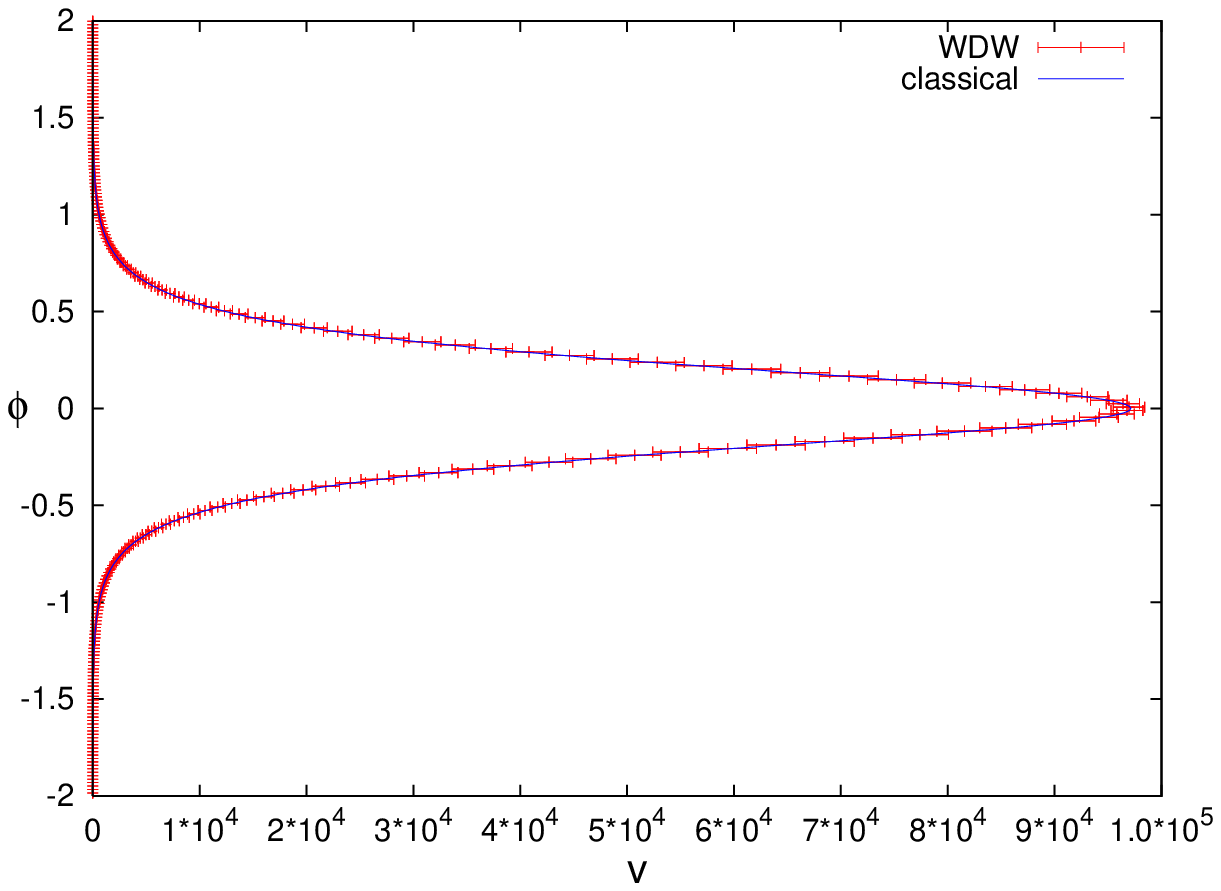}
\caption{$a)$ The absolute value of the \WDW wave function. The initial
state at `time $\phi_o$ was constructed using (\ref{sc}) and is
sharply peaked at $p^\star_\phi$, and $v^\star|_{\phi_o}$. It
remains sharply peaked on the classical trajectory with $p_\phi
=p_\phi^\star$, passing through $v^\star$ at $\phi=\phi_o$. For
clarity of visualization, only the values of $|\Psi|$ greater than
$10^{-4}$ are shown. Being a physical state, $\Psi$ is symmetric under
$v \rightarrow -v$. In this simulation, the parameters were:
$p_{\phi}^{\star} = 5000 $, and $\Delta p_\phi/p_\phi^{\star} = 0.02$.
$b)$ The expectation values (and dispersions) of
     ${|\hat{v}|_{\phi}}$ are plotted for the WDW wave function
     and compared with the
     classical trajectory. The \WDW wave function follows the
     classical trajectory into the big-bang and big-crunch
     singularities.
}
    \label{fig:wdw1}
  \end{center}
\end{figure}

Using the physical Hilbert space and this complete set of Dirac
observables we can now introduce semi-classical states and study
their evolution. Let us fix an `instant of time' $\phi=\phi_o$ and
construct a semi-classical state which is peaked at $p_\phi =
p_\phi^\star$ and $|v|_{\phi_o} = v^\star$. Since we would like
the peak to be at a point that represents a large classical
universe, we are led to choose $v^\star \gg 1$ and \,
$p_\phi^\star \gg \hbar$\, (in the natural classical units
$c$=$G$=1).  The second condition is necessary to ensure that the
universe expands out to a size much larger than the Planck scale.
At `time' $\phi=\phi_o$, consider the state
\be \label{sc} \ul{\Psi}(v,\phi_o) = \int_{0}^\infty \dd k\,\,
\t{\Psi}(k)\,\, \ub{e}_k(v)\, e^{i\omega\,(\phi_o-\phi^\star)},
\quad {\rm where}\,\, \t{\Psi}(k) =
e^{-\f{(k-k^\star)^2}{2\sigma^2}}\, . \ee
Here
\be k^\star = \sqrt{3/16\pi G \hbar^2}\,\, p_\phi^\star \quad {\rm
and} \quad \phi^\star = \phi_o + \sqrt{\f{3}{16\pi G}}\, \cosh^{-1}
\big[ (\f{3K^2(p_\phi^{\star})^2}{\lo^2 G \hbar^2
\gamma^2})^{1/2}\, \, (v^\star)^{-2/3} \big]\, . \ee
In the spatially flat case, the eigenfunctions $e_k(v)$ were just
plane waves \cite{aps3} and one could evaluate this integral
analytically. The modified Bessel functions are much more
complicated. Therefore, in the closed model we have to use
numerical methods. They show that this state is sharply peaked at
values $v^\star$, $p_\phi^\star$ of our Dirac observables (see
Fig. \ref{fig:wdw1}a).

We can now ask for the evolution of this state. Does it remain
peaked at the classical trajectory defined by $p_\phi =
p_\phi^\star$ and passing through $v=v^\star$ at $\phi =\phi_o$?
This question is easy to answer because (\ref{sol2}) implies that
the (positive frequency) solution $\ul{\Psi}(v,\phi)$ to
(\ref{wdw1}) defined by the initial data (\ref{sc}) is obtained
simply by replacing $\phi_o$ by $\phi$ in (\ref{sc})! Since
$\sigma$, the measure of dispersion in (\ref{sc}), does not depend
on $\phi$, it follows that $\ul{\Psi}(v,\phi)$ continues to be
peaked at a trajectory
which is precisely the classical solution of interest. This is
just what one would hope during the epoch in which the universe is
large. However, the property holds also in the Planck regime and
the semi-classical state simply follows the classical trajectory
into the big-bang and big crunch singularities. (See Figs.
\ref{fig:wdw1}a and \ref{fig:wdw1}b.) In this sense, the \WDW
evolution does not resolve the classical singularity.

We will show in the next three sections that the situation is very
different in LQC. This can occur because the \WDW equation is a
good approximation to the discrete equation only for large $v$.
Furthermore, as discussed in \cite{aps3}, the approximation is not
uniform but depends on the state: in arriving at the \WDW equation
from LQC we had to neglect $\Psi$ dependent terms of the form
$O(v^{n-3}\f{\dd^n\Psi}{\dd v^n})$ for $n\ge 3$. For
semi-classical states considered above, this implies that the
approximation is excellent for $v \gg k^\star$ but becomes
inadequate when the peak of the wave function lies at a value of
$v$ comparable to $k^\star$. Then, the LQC evolution departs
sharply from the \WDW evolution. We will find that, rather than
following the classical trajectory into the big bang singularity,
the peak of the LQC wave function now exhibits a bounce. Since
large values of $k^\star$ are classically preferred, the value of
$v$ at the bounce can be quite large. However, as in the $k = 0$
models \cite{aps3}, we will find that the matter density at the bounce
point is comparable to the Planck density, independently of the
precise value of $k^\star$ so long as $p_\phi^\star \gg \hbar$.

\section{LOOP QUANTUM COSMOLOGY: ANALYTICAL ISSUES}
\label{s4}

Since generic solutions in the closed, $k$=1 cosmologies undergo a
recollapse in classical general relativity, the scale factor can
not serve as a global time variable. This fact has been used as a
criticism of the implicit use of the scale factor as time in the
older LQC literature (see, e.g., \cite{gu}). However, if matter
sources \emph{include} a massless scalar field $\phi$, that field
\emph{is} monotonic and single valued in all classical solutions.
Therefore it can be chosen as the `internal' time variable with
respect to which the scale factor and other fields evolve.
Furthermore, as discussed in section \ref{s2}, for the model
considered in this paper the Hamiltonian constraint (\ref{hc2}) of
LQC has the same form as the wave equation in a static space-time,
with $\phi$ playing the role of time and $\Theta$ of the elliptic
spatial operator plus a static potential. Therefore as in the
$k$=0 case $\phi$ can be regarded as \emph{emergent time} also in
the quantum theory. This choice is
free of the criticisms mentioned above.%
\footnote{Furthermore, as emphasized in \cite{aps2,aps3}, while
the availability of a global time variable simplifies the
constructions considerably and makes physical interpretation
transparent, it is not essential. Using the group averaging
procedure \cite{dm}, for example, the physical sector of the
theory can be constructed even when a global intrinsic time does
not exist. Furthermore, in simple examples \cite{lr} physical
interpretation can be aided by the introduction of a suitable
\emph{local} notion of time which exists, e.g., if the scalar
field comes with an inflationary potential.}
In this section we will construct the physical sector of LQC by
exploiting this fact. As mentioned in section \ref{s1}, we will
find that quantum geometry effects resolve both the big bang and
the big crunch singularities, leading to a cyclic quantum
universe. Still the scalar field will continue to be a viable
emergent time \emph{globally}.

\subsection{General solution to the LQC Hamiltonian constraint}
\label{s4.1}

As in the spatially flat $k$=0 case, our kinematical Hilbert space
will be: $\Hk := L^2(\R_{\rm Bohr}, B(v)\dd\mu_{\rm Bohr}) \otimes
L^2(\R, \dd \phi)$. Since $\phi$ is to be thought of as `time' and
$v$ as the genuine, physical degree of freedom which evolves with
respect to this `time', we have chosen the standard Schr\"odinger
representation for $\phi$ but the `polymer representation' for $v$
to correctly incorporate the quantum geometry effects. This is a
conservative approach in that the results will directly reveal the
manifestations of quantum geometry. Had we chosen a non-standard
representation for the scalar field, these effects would have been
mixed with those arising from an unusual representation of `time
evolution' and, furthermore, comparison with the \WDW theory would
have become more complicated.

The form of the LQC Hamiltonian constraint (\ref{hc2}) is the same
as that of the \WDW constraint (\ref{wdw1}). Properties of
$\Theta_o$ analyzed in \cite{aps3} immediately imply that $\Theta$
is again a positive, symmetric operator on $L^2(\R_{\rm Bohr},
B(v)\dd\mu_{\rm Bohr})$, whence it admits a self-adjoint
(Friedrich) extension. (For precise domains, see \cite{warsaw}.)
The main difference is that while the \WDW $\ul{\Theta}$ is a
differential operator, the LQC $\Theta$ is a \emph{difference}
operator. This gives rise to certain technically important
distinctions. For, now the space of physical states
---i.e. of appropriate solutions to the constraint equation--- is
naturally divided into sectors each of which is preserved by the
`evolution' and by the action of our Dirac observables. Thus,
there is super-selection. Let $\La_{|\varepsilon|}$ denote the
`lattice' of points $\{|\varepsilon|+4n,\, n\in \Z\}$ on the
$v$-axis, $\La_{-|\varepsilon|}$ the `lattice' of points
$\{-|\varepsilon|+4n,\, n\in \Z\}$ and let $\La_{\varepsilon} =
\La_{|\varepsilon|} \cup \La_{-|\varepsilon|}$ where as usual $\Z$
denotes the set of integers. Let $\H_{|\varepsilon|}^{\rm
grav},\H_{-|\varepsilon|}^{\rm grav}$ and $\H_{\varepsilon}^{\rm
grav}$ denote the subspaces of $L^2(\R_{\rm Bohr}, B(v)
\dd\mu_{\rm Bohr})$ with states whose support is restricted to
lattices $\La_{|\varepsilon|}, \La_{-|\varepsilon|}$ and
$\La_\varepsilon$. Each of these three subspaces is mapped to
itself by $\Theta$ which is self-adjoint and positive definite on
all three Hilbert spaces. However, for reasons explained in detail
in \cite{aps3}, our physical states will be invariant under the
orientation reversing operator $\Pi$ of (\ref{pi}). Thus, we are
primarily interested in the symmetric subspace of
$\H_\varepsilon^{\rm grav}$; the other two Hilbert spaces will be
useful only in the intermediate stages of our discussion.

Our first task is to explore properties of the operator $\Theta$.
Since it is self-adjoint and positive definite, its spectrum is
non-negative. Therefore as in the \WDW theory we will denote its
eigenvalues by $\omega^2$. On each Hilbert space
$\H_{\pm|\varepsilon|}^{\rm grav}$, we can solve for the
eigenvalue equation $\Theta\, \psi_\omega(v) = \omega^2\,
\psi_\omega (v)$, i.e.,
\ba  \label{eq:eigen} &&C^+(v)\, \psi_\omega(v+4) +
C^o(v)\,\psi_\omega(v) + C^-(v)\, \psi_\omega(v-4) \nonumber\\
&+& \f{\pi G}{2} \bigg[3K (\sin^2(\f{\mb\lo}{2}) -
\f{\mb^2\lo^2}{4})\, |v|\, - \, \f{1}{3}\, \lo^2\g^2\,\,
\big|\f{v}{K}\big|^{\fs{1}{3}}\bigg] \,\,\, \bigg[\big|\,
 |v-1|- |v+1|\,\big|\bigg]\, \psi_\omega(v)\nonumber \\
&=& \omega^2 B(v) \psi_\omega(v)\, . \ea
Since this equation has the form of a second order recursion
relation, as in the $k$=0 case one might expect a 2-fold
degeneracy. However, there is an important subtlety. Let us
consider the asymptotic regime $v \gg 1$. Then, each $\psi_\omega$
approaches a solution to the \WDW equation and as we saw in
section \ref{s3.1} there is only one linearly independent solution
which does not diverge for large $v$. The form of the inner
product on $\H_\varepsilon^{\rm grav}$ now implies that, as in the
\WDW theory, the degeneracy is removed; only one of the two
linearly independent solutions can belong to the spectral family
of $\Theta$. Furthermore, numerical calculations show that this
solution, which decays exponentially for $v \gg 1$, will in
general diverge in the other asymptotic limit $-v \gg 1$. It is
only for some \emph{discrete} values $\omega_n$ of $\omega$ that
one obtains solutions which do not diverge in either asymptotic
limits. (This phenomenon was already noted in \cite{NPV} in a
simpler model without a scalar field but with a negative
cosmological constant.) Since these solutions decay exponentially
in both limits, they are normalizable in $\H_\varepsilon^{\rm
grav}$. Thus, in contrast to the $k$=0 case, on each superselected
sector, the spectrum of $\Theta$ is \emph{discrete and each of its
eigenvalues is non-degenerate}.%
\footnote{See \cite{warsaw} for an analytical proof which does not
rely on numerical results. Recall that in the \WDW theory the
spectrum is non-degenerate but \emph{continuous}. This difference
arises because in the \WDW theory the eigenfunctions oscillate
more and more wildly as $v$ goes to zero and thus fail to be
normalizable in the \WDW Hilbert space $L^2(\R, (K/v)\dd v)$.
Because of the discreteness of lattices $\La_{\pm|\varepsilon|}$,
nothing special happens `near' $v=0$ in LQC.}
We will denote the normalized eigenfunctions in
$\H_{|\varepsilon|}^{\rm grav}$ by $e_{n}(v) $:
\be \Theta\, e_n(v)= \omega_n^2\, e_n(v)\quad {\rm and}\quad
\ip{e_m}{e_n} = \delta_{m,n}\, . \ee
Finally, physical states will be built from eigenfunctions
$e^{(s)}_n$ which are symmetric under orientation reversal. Since
$\Theta$ commutes with the orientation reversal operator $\Pi$,
$e_n(-v) = \Pi e_n(v)$ is an eigenfunction of $\Theta$ with the
same eigenvalue $\omega_n^2$ as $e_n(v)$, but belongs to
$\H_{-|\varepsilon|}^{\rm grav}$ rather than
$\H_{|\varepsilon|}^{\rm grav}$. Therefore,
\be e^{(s)}_n = \f{1}{\sqrt{2}}\, (e_n(v) + e_n(-v))\, , \ee
also has eigenvalue $\omega^2_n$, but belongs to
$\H_\varepsilon^{\rm grav}$.

We can now write down the general symmetric solution to the
quantum constraint (\ref{hc2}) with initial data in
$\H_{\varepsilon}^{\rm grav}$ :
\be \label{sol3}\Psi(v,\phi) = \sum_n \,\, [\t\Psi_n^+ \,e^{(s)}_n
(v)\, e^{i\omega_n\phi} + \t\Psi_n^- \,\bar{e}^{(s)}_n(v)\,
e^{-i\omega_n\phi}]\ee
where $\t{\Psi}_n^\pm$ are square-summable. As in the \WDW theory,
if $\t\Psi_n^-$ vanishes, we will say that the solution is of
positive frequency and if $\t\Psi_n^+$ vanishes we will say it is
of negative frequency. Thus, every solution to (\ref{hc2}) admits
a natural positive and negative frequency decomposition. The
positive (respectively negative) frequency solutions satisfy a
Schr\"odinger type first order differential equation in $\phi$:
\be \label{eq:fund_eq} \mp i\f{\p\Psi_\pm}{\p\phi} = \sqrt{\Theta}
\Psi_\pm \ee
with a Hamiltonian $\sqrt{\Theta}$ (which is non-local in $v$).
Therefore the solutions with initial datum $\Psi(v, \phi_o) =
f_\pm(v)$ are given by:
\be \Psi_\pm(v,\phi) \,=\, e^{\pm
i\sqrt{\Theta}\,\,(\phi-\phi_o)}\, f_\pm(v,\phi)\, .\ee

To summarize, the overall structure is analogous to that in the
spatially open, $k$=0 case. A key difference is that the spectrum
of $\Theta$ is discrete on each of the three Hilbert spaces,
$\H_{\pm|\varepsilon|}^{\rm grav}$ and $\H_{\varepsilon}^{\rm
grav}$. In particular, while in the $k$=0 case all eigenfunctions
have an oscillatory asymptotic behavior, now they all decay
exponentially for sufficiently large $|v|$. This difference neatly
encodes in the quantum theory the key qualitative difference
between the two models in the classical theory: In the $k$=0 case
any one classical solution is either ever expanding or ever
contracting, while in the $k$=1 case each solution expands to a
maximum volume and then recollapses.

\subsection{The Physical sector}
\label{s4.2}

We will now summarize the mathematical structure of the physical
sector of the theory. The construction is entirely analogous to
that in the $k$=0 case since the spectrum of the full constraint
operator $\p_\phi^2 + \Theta$ is still continuous (because of the
$\p_\phi^2$ part). Therefore we will only state the final results.

The sector of the physical Hilbert space $\Hp^\varepsilon$
labeled by $\varepsilon \in [0,\, 2 ]$ consists of positive
frequency solutions $\Psi(v,\phi)$ to (\ref{eq:fund_eq}) with
initial data $\Psi (v, \phi_o)$ in the symmetric sector of
$\H^\varepsilon_{\rm grav}$. Eq. (\ref{sol3}) implies that they
admit an explicit expansion
\be \label{eq:psi-int} \Psi(v,\phi) = \sum_n\, \t\Psi_n \,\,
e^{(s)}_n(v)\,\, e^{i\omega_n\phi}\,\, , \ee
where, we have suppressed the superscript + because from now on we
will only work with positive frequency solutions. The physical
inner product is given by
\be \label{ip2} \langle\Psi_1|\Psi_2\rangle_\varepsilon\,\, =
\sum_{v \in \{\pm|\varepsilon| + 4n;\, n\in \Z\}}\,\, B(v)\,
\bar{\Psi}_1(v,\phi_o)\,\, \Psi_2(v, \phi_o) \ee
for any $\phi_o$. The action of the Dirac observables is
independent of $\varepsilon$, and has the same form as in the \WDW
theory:
\be \label{dirac4}{|\h{v}|_{\phi_o}}\, \Psi(v,\phi) =
e^{i\sqrt{{\Theta}}\,\,(\phi-\phi_o)}\,\, |v|\,
\,\Psi(v,\phi_o),\quad {\rm and} \quad \hat{p}_\phi \Psi(v,\phi) =
- \, i\hbar\, \f{\p \Psi(v,\phi)}{\p\phi} \, . \ee
The kinematical Hilbert space $\Hk$ is non-separable but, because
of super-selection, each physical sector $\Hp^\varepsilon$ is
separable. Eigenvalues of the Dirac observable
${|\h{v}|_{\phi_o}}$ constitute a discrete subset of the real line
in each sector. The set of these eigenvalues in different sectors
is distinct. Therefore which sector  actually occurs is a question
that can be in principle answered experimentally, provided one has
access to microscopic measurements which can distinguish between
values of the scale factor which differ by about a Planck length.
This will not be feasible in the foreseeable future. Of greater
practical interest are the coarse-grained measurements, where the
coarse graining occurs at significantly greater scales. For these
measurements, different sectors would be indistinguishable and one
could work with any one.

\section{LOOP QUANTUM COSMOLOGY: NUMERICAL ISSUES}
\label{s5}

As we saw in section \ref{s4} physical states can be readily
constructed from eigenfunctions of the difference operator
$\Theta$. In the first part of this section, we study properties
of these eigenfunctions. We show that $\Theta$ \emph{admits only
normalizable eigenfunctions with discrete eigenvalues} and
numerically construct an orthonormal basis. In the second part
we use this basis to construct and analyze physical semi-classical
states.

\subsection{Spectrum of $\Theta$}
\label{s5.1}

Eigenfunctions of $\Theta$ are solutions to the difference
equation \eqref{eq:eigen}. Since its coefficients are real, any
eigenfunction can be expressed as a complex linear combination of
real eigenfunctions. Therefore, it will suffice to restrict
ourselves to real eigenfunctions.

Consider a generic lattice $\La_{\vep}$, i.e., a lattice where
$\varepsilon$ does not equal 0 or 2. Since $\La_{\vep} = \La_{|\vep|} \cup
\La_{-|\vep|}$, to obtain an eigenfunction which is symmetric
under orientation reversal $v \rightarrow -v$ it suffices to solve
the eigenvalue equation just on $\La_{|\vep|}$ and then reflect
it. Because $\Theta$ is invariant under orientation reversal, the
reflected function is automatically an eigenfunction with the same
eigenvalue, but supported on
$\La_{-|\vep|}$. Furthermore, if the original eigenfunction is
normalizable on $\La_{|\vep|}$, the reflected one is normalizable
on $\La_{-|\vep|}$ whence the sum is a symmetric, normalizable
eigenfunction on $\La_{\vep}$. Therefore, in what follows,
\emph{for a generic $\vep$ we will restrict ourselves to
$\La_{|\vep|}$ and examine all eigenfunctions.} Since lattices
with $\vep = 0$ or $2$ are symmetric under reflection, \emph{on
these exceptional lattices we will restrict ourselves only to
symmetric eigenfunctions.}

Every such eigenfunction $\psi_{\omega} (v)$ is uniquely
determined by its `initial' values $\psi_{\omega}(\vep+4n)$,
$\psi_{\omega}(\vep+4(n+1))$ for some integer $n$.%
\footnote{{\setstretch{0.1} If $\varepsilon\neq 2$, the eigenvalue
equation \eqref{eq:eigen} recursively determines $\psi_{\omega}(v)$
on the entire $\La_{+|\varepsilon|}$. If $\varepsilon=2$ the
coefficients $C^-(2)$ and $C^+(-2)$ in Eq. \eqref{eq:eigen}
vanish. Therefore, we can calculate $\psi_{\omega}(v)$ only on half
of the $v$ axis. However, values of $\psi_{\omega}(v)$ on the other
half are determined by the symmetry condition.}}
For later convenience, we note that the initial data can be
represented by a pair of real parameters $b\in\mathbb{R}$ and
$\theta\in[0,\pi]$
\begin{subequations}\label{eq:eig-ID}\begin{align}
  \psi_{\omega}(\vep+4n)\ &=\ b\cos(\theta) \ , &
  \psi_{\omega}(\vep+4(n+1))\ &=\ b\sin(\theta) \ .
    \tag{\ref{eq:eig-ID}}
\end{align}\end{subequations}
Thus on generic lattices, for any $\omega$, the eigenspace is
\emph{2-dimensional} and the degeneracy is parameterized by
$b,\theta$. On the exceptional lattices, the symmetry requirement
imposes an additional constraint which determines
$\psi_{\omega}(\vep+4)$ as function of $\psi_{\omega}(\vep)$.
Therefore on these lattices the eigenspace is only \emph{one
dimensional}.

\begin{figure}[tbh!]
  \begin{center}
    $a)$\hspace{8cm}$b)$
    \includegraphics[width=3.2in,angle=0]{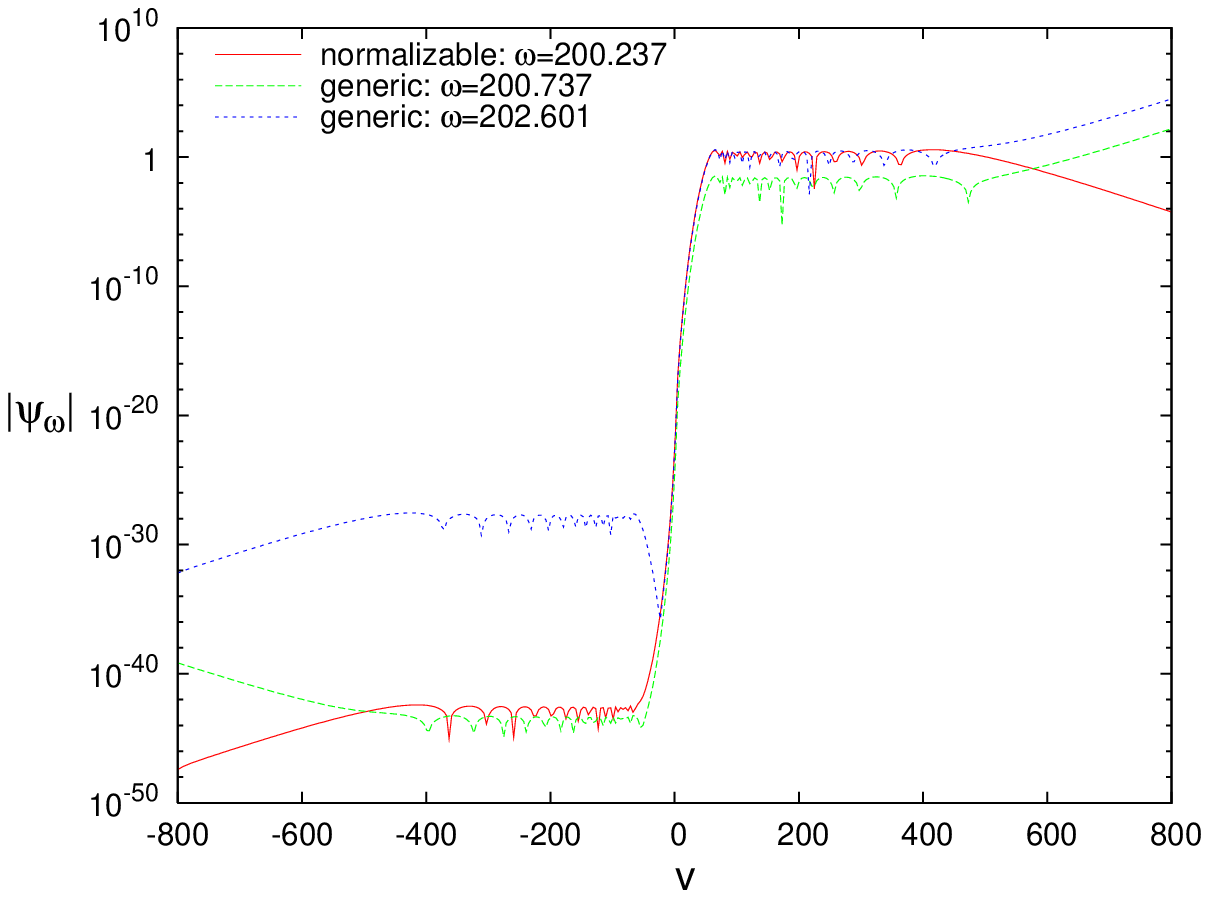}
    \includegraphics[width=3.2in,angle=0]{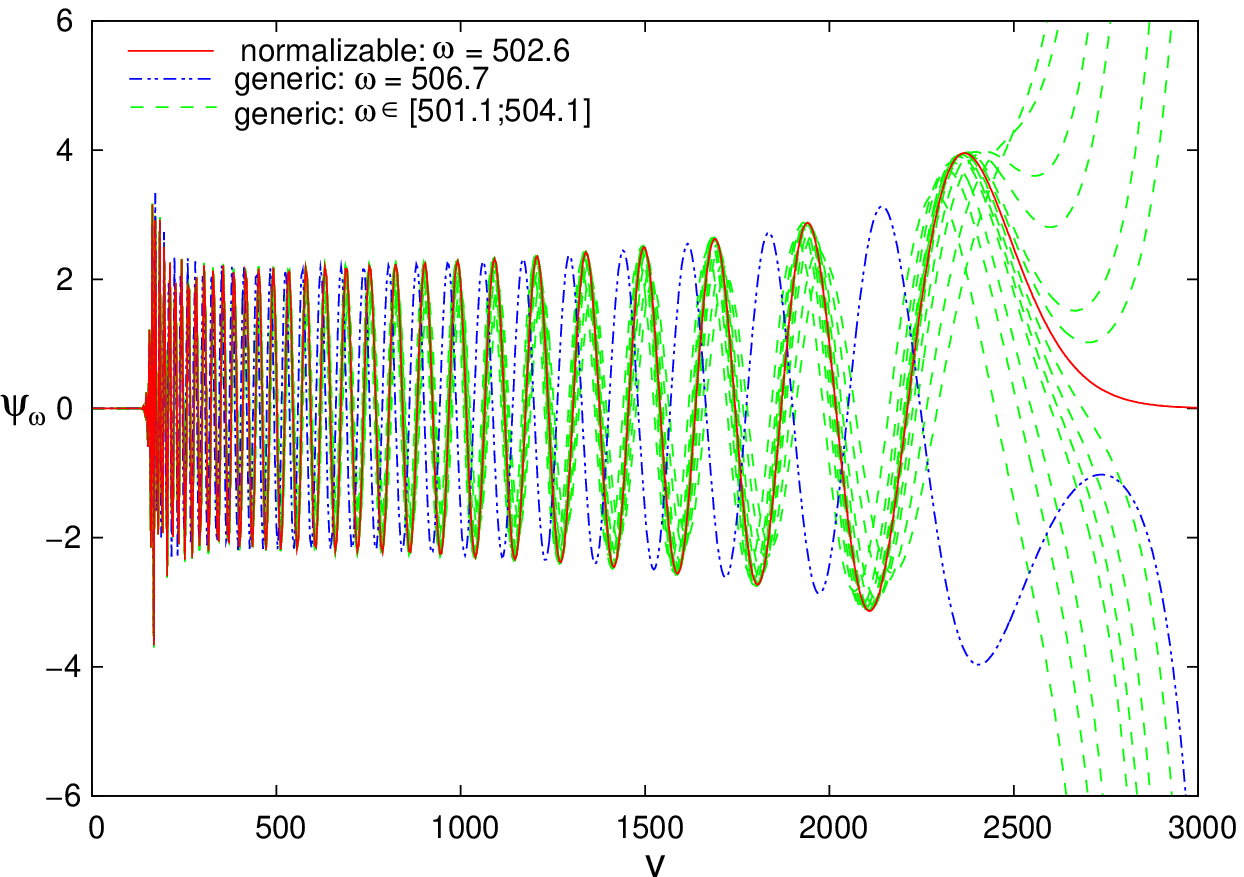}
    \caption{Examples of normalizable and generic eigenfunctions of
$\Theta$. The plot $a)$ presents eigenfunctions supported on the
lattice with $\vep=0.5$. Since the scale on the y-axis is
logarithmic, it is clear that in the asymptotic region generic
eigenfunctions diverge exponentially while the normalizable ones
decay exponentially. Plot $b)$ shows eigenfunctions supported on
the $\vep=0$ lattice. Since the scale on the y-axis is linear this
plot brings out the detailed behavior of eigenfunctions in the
region where $|\psi_\omega|$ is small.}
    \label{fig:eig-p-log}
  \end{center}
\end{figure}

The generic behavior of eigenfunctions $\psi_\omega(v)$ is shown
in Fig \ref{fig:eig-p-log}. One can distinguish three regions on
the positive (or negative) $v$ axis in each of which the
eigenfunction $\psi_\omega(v)$ has a qualitatively different
behavior.

\begin{enumerate}[(i) ]
  \item Genuine quantum region: $|v|\lesssim v_b$ in which
 $|\psi_{\omega}|$ grows/decays exponentially. For $\varepsilon=0$
 or $2$, $|\psi_{\omega}|$ always grows with $|v|$. $v_b$ turns out
 to be proportional to $\omega$.
  \item `Semi-classical' region: $v_b \lesssim |v|\lesssim v_r$ in
  which $\psi_\omega$ oscillates. $v_r$ turns out to be
  proportional to $\omega^{3/2}$  (and approximately equals the maximal
  $|v|$ of a classical universe of momentum $p_{\phi}=\hbar\omega$).
 \item Classically forbidden region:  $|v| \gtrsim v_r$, where
 $\psi_\omega$ grows/decays exponentially as $|v|$ increases.
    \label{it:class-forb}
\end{enumerate}
(However, numerical simulations show that the distinction between
first two regions gets blurred for eigenfunctions corresponding to
$\omega \lesssim 5$.) Note that for a generic lattice, the
eigenfunction may decay in one asymptotic region (say, $v
\rightarrow \infty$) but grow in the other ($v \rightarrow
-\infty$). Eigenfunctions growing on either side fail to be
normalizable, whereas the ones decaying for both signs of $v$ are
normalizable. While eigenfunctions exist for any
$\omega$, they decay on both sides only for certain discrete
values of $\omega$. Consequently the spectrum of $\Theta$ is
discrete and the normalizable eigenfunctions form a set of zero
measure in the space of all eigenfunctions. We will now describe
the search algorithm to find them. We begin with the simpler case
of $\vep=0$ or $2$ lattices and then discuss the more subtle case
of generic lattices.

Fix $\varepsilon=0$ or $2$. Because of symmetry, it suffices to
analyze the behavior of $\psi_{\omega}(v)$ just for $v>0$. Let us
focus on an interval $W=[\omega_1,\omega_2]$ of frequencies and
fix a point $v'\in \La_{\vep}$ such that $v'\gg v_r(\omega_2)$.
Eq. \eqref{eq:eigen} implies that the value $\psi_{\omega}(v')$ is
a continuous function of $\psi_{\omega}(v)|_{v=\vep}$ and
$\omega$. One can fix the initial value $\psi_{\omega}(\vep)$ to
$1$, thus leaving the dependence only on $\omega$. Numerical
inspection shows that $\psi_{\omega}(v)$ changes sign
quasi-periodically as $\omega$ increases (see the right plot of
Fig. \ref{fig:eig-p-log}). Let us take one of the (possibly many)
values $\omega_{n,v'}$ such that $\psi_{\omega_{n,v'}}(v)|_{v=v'}
= 0$. The limits
\begin{equation}
  \omega_n = \lim_{v'\to\infty} \omega_{n,v'} \
\end{equation}
are the only eigenvalues corresponding to normalizable
eigenfunctions. In practice the values $\omega_{n,v'}$ for
$v'\approx 1.3v_r$ approximate the limiting value $\omega_n$ with
precision $10^{-16}$.

In actual calculations the following algorithm was applied:
\begin{enumerate}[(i)]
  \item First we consider a set of frequency values $\omega_i$ uniformly
    distributed within the interval $[0,\omega_{\textrm{max}}]$. For each
    fixed $v'$, the separation $\omega_{i+1}-\omega_i$ was chosen to be
    much smaller than separation between values of $\omega$ at
    which $\psi_{\omega}(v')$ vanishes. (In practice the separations
    turned out to be greater than $1$ for $\omega<3\times 10^5$.)
  \item Whenever the change of sign between $\psi_{\omega_{i+1}}(v')$
    and $\psi_{\omega_{i}}(v')$ was detected, the value $\omega_{n,v'}$
    corresponding to the root of $\psi_{\omega}(v')$ was found via
    bisection method.
\end{enumerate}
For this scan, $\omega_{\textrm{max}}$  was chosen as $3\times
10^5$. The first of the two steps in the algorithm ensures that
all eigenfunctions in the chosen interval have been found.

Let us now consider a generic lattice. Now there are two factors
which complicate the task of finding normalizable eigenfunctions.
First, the eigenspaces are \emph{two} dimensional, parameterized
by $b, \theta$ as in \eqref{eq:eig-ID}. Therefore, even if we fix
the normalization freedom by setting $b=1$, for each frequency
$\omega$ we have a \emph{1-parameter family} of eigenfunctions,
labeled by $\theta$ (rather than a single eigenfunction as on
$\vep=0$ or $2$ cases). Secondly, since the desired eigenfunctions
do not have to be symmetric on $\La_{|\vep|}$, now we have to
analyze the behavior of $\psi_{\omega}$ when $|v| \gg v_r$
\emph{separately for positive and negative values} of $v$. We therefore
modified the algorithm specified for the $\vep=0$ or $2$ lattices
as follows.

First, keeping $\omega$ fixed and varying $\theta$ instead of
$\omega$ in the above procedure, we searched for
$\psi_{\omega}(v)$ which decays on the \emph{negative $v$ side}.
For this we probed the domain of $\theta$ in $100$ points
uniformly distributed within $[0,\pi]$ and, whenever sign of
$\psi_{\omega}(v')$ changed, we narrowed the choices of $\theta$
using bisection. The analysis shows that for each $\omega$ there
exists {\it unique} $\theta$ such that $\psi_{\omega}$ satisfies
this condition. With $\theta$ so determined, we again have a
1-dimensional eigenspace $\psi_\omega(v)$ for each $\omega$.
Therefore, we could now apply the procedure used for $\vep=0$ or
$2$ to look for eigenfunctions which decay \emph{on the positive
side}. Thus the two complications so to say compensate one
another: while there is an additional, 1-dimensional freedom
(labeled by $\theta$) in the choice of eigenfunctions, since the
requirement that the eigenfunctions decay on the negative $v$ side
is now decoupled from the requirement that they decay on the
positive side, we have an additional constraint (which determines
$\theta$ for any given $\omega$). Therefore, the only modification
to the algorithm used in the $\vep=0$ or $2$ cases was to first
determine $\theta$.

To gain insight into the qualitative features of eigenfunctions
and eigenvalues, this method was first applied to $\omega\in
[0,50]$ by choosing $\La_{+|\varepsilon|}$. Normalized
eigenfunctions decay exponentially as $|v|$ increases for both
positive and negative $v$. It was found that these eigenfunctions
generically have an additional feature: in the `semi-classical'
region described above, their values on the positive or negative
sides of the $v$ axis are suppressed relative to their values on
the other half, the difference in the amplitudes growing
exponentially with $\omega$. Since we will be primarily interested
in large $\omega$, if the initial data are specified on the side
where $\psi_{\omega}$ is large, the suppression on the other side
enhances the numerical errors making the results unreliable. For
these cases, the search of normalizable eigenfunctions was
performed again, now starting from the side where $\psi_\omega$ is
small.

After implementing all these precautions to control errors, a much
more exhaustive search for eigenvalues and eigenfunctions was
made. The results can be summarized as follows.

\begin{figure}[]
  \begin{center}
    $a)$\hspace{8cm}$b)$
    \includegraphics[width=3.2in,angle=0]{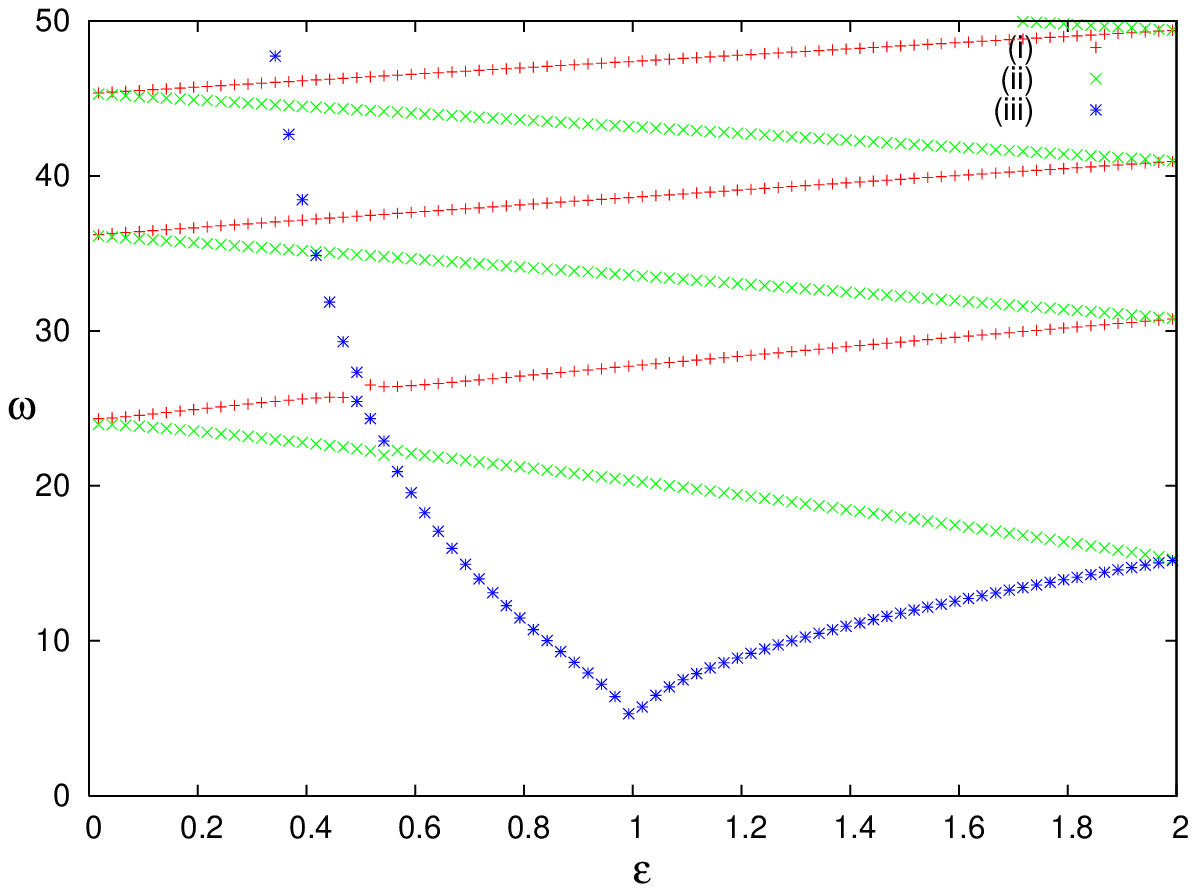}
    \includegraphics[width=3.2in,angle=0]{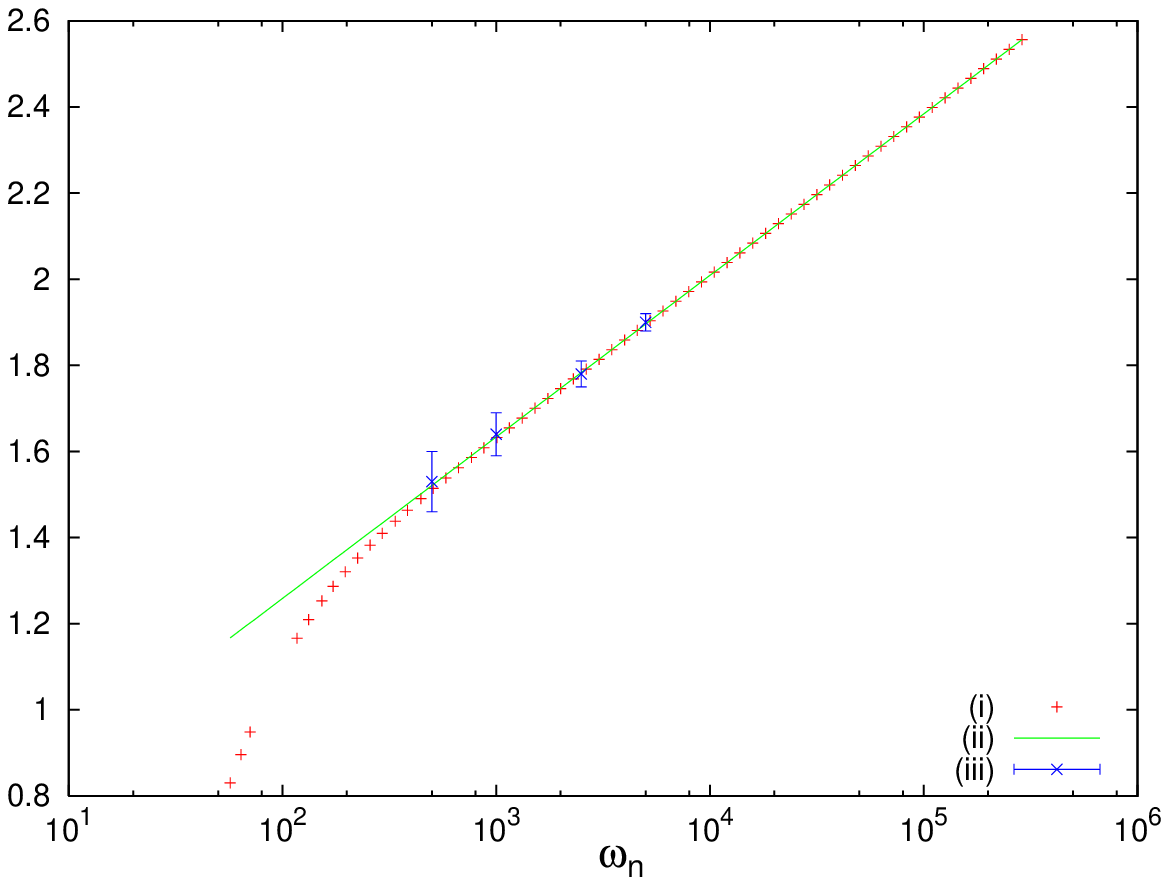}
    \caption{$a)$ Distribution of lowest ($\omega<50$) eigenvalues
      (corresponding to normalizable eigenfunctions) for different $\vep$.
      Eigenvalues are divided into three classes according to
      the behavior of eigenfunctions, which are i) suppressed for $v<0$,
      ii) suppressed for $v>0$, and iii) large only near $v=0$.
      \,\,
      $b)$ Rescaled spectral density $2\pi/(\omega_{n}-\omega_{n-1})$ (i)
      is compared with periods $T_\phi(\omega_n)$ in $\phi$ (iii) of
      expectation value of $|\hat{v}|_{\phi}$ of the coherent states
      evolved from the \WDW initial data (see section \ref{s5.2}). The
      solid line (ii) shows the large $\omega$ limit of the rescaled
      spectral density.}
    \label{fig:low-omega}
  \end{center}
\end{figure}

\begin{enumerate}[(i) ]
  \item The spectrum of $\Theta$ is discrete for all lattices.
    Each of its eigenvalues is non-degenerate. Normalizable
    eigenfunctions decay exponentially as $|v|$ tends to infinity.
  \item For generic $\vep$, the restrictions to sublattices
  $\La_{\pm|\varepsilon|}$ of eigenfunctions $\psi_n(v)$ are
    in general strongly suppressed for one sign of $v$. The side
    on which suppression occurs depends on parity of $n$.
  \item The density of eigenvalues has following features. As shown in
    Fig. \ref{fig:low-omega}a it is roughly independent of $\vep$
    except that for generic lattices it is twice as large as that
    for exceptional ones. It slowly grows with $\omega$ (see Fig.
    \ref{fig:low-omega}b). The best fit to the data in
    Fig. \ref{fig:low-omega}b leads to the behavior in large $\omega$
    limit as:
    \be \label{eq:dw-est1} \omega_{n}-\omega_{n-1} \rightarrow
    \f{1}{\alpha \ln|\omega_n/\beta|}\ee
     with $\alpha\ =\ [0.0259272\pm (5 \times 10^{-7})] G^{-1/2}$
     and $\beta\ =\ [0.04412\pm (1 \times 10^{-5})] G^{1/2}.$
  \item A state $\Psi(v)$ which is sharply peaked at some $\omega^\star$
  can be well-approximated by a linear combination only of eigenfunctions
  with eigenvalues lying in a small compact interval around $\omega^\star$.
  If $\omega^\star$ is very large the estimate given in (iii) above implies
  that the distribution of $\omega_n$ is approximately uniform.
  Consequently, the wave function will be approximately periodic in
  $\phi$ with a period
    \begin{equation}
      T_{\phi}(\omega_n) \approx \frac{2\pi}{\omega_n-\omega_{n-1}} \ .
    \end{equation}
    For an LQC physical state obtained from evolution of the initial
    data at $\phi=\phi_o$ corresponding to a \WDW coherent state (see
    section \ref{s5.2}), the period of expectation value $\langle
    |\hat{v}|_{\phi} \rangle$ turns out to be in good agreement with
    $T_{\phi}$. This provides an independent check on our numerical analysis.
\end{enumerate}


\subsection{Evaluation of Semi-classical State and Observables}
\label{s5.2}

The numerical method presented in previous subsection allow us to
find all the eigenstates of $\Theta$ which span the physical
Hilbert space. To obtain the normalized eigenbasis $e^{(s)}_n$
(which is symmetric under orientation reversal) we first note that
in the expression of the norm it suffices to evaluate just a
finite sum
\begin{equation}
   \| \psi_n \|^2_{\varepsilon}\
   =\ \sum_{v\in\{\pm\varepsilon+n;\,-N<n<N\}} B(v)\psi^2_n(v) , \quad
   \,\, \mathrm{where} \,\,  N > \f{v_r}{4}.
\end{equation}
since $\psi_n(v)$ decay exponentially for $|v| > v_r \propto
\omega^{\frac{3}{2}}$. With the basis at hand, one can construct
physical states which are semiclassical at late times. As in $k=0$
case \cite{aps2}, this can be done in two different ways:
\begin{enumerate}[(i) ]
  \item Direct evaluation of the wave function using
    \eqref{eq:psi-int}, and
  \item Evolution in $\phi$ of the initial data specified at
    $\phi=\phi_o$ using Eq. \eqref{hc2}. (The initial data can be
    chosen to be the same as that of a semi-classical solution to
    the \WDW equation at some late `time' (see section \ref{s3})).
    \label{it:evol}
\end{enumerate}
The direct evaluation of the integral expression
\eqref{eq:psi-int} already provides the full LQC solution.
However, we also used this expression to obtain just the initial
data at $\phi=\phi_o$ and then evolved this data using
\eqref{hc2}. Agreement between the two solutions provides an
independent check on numerics. Finally, we also evolved the
initial data extracted from a \WDW coherent state and used the
resulting LQC solution thereby obtaining an independent check on
completeness of the eigenbasis constructed in section \ref{s5.1}.
The rest of this section provides the relevant details of the
numerical implementation of this procedure.

 In order to evaluate the integral solution, we need the
spectral profile $\tilde{\Psi}_n$ used in the expansion
(\ref{sol3}). Since we are primarily interested in states which
are semiclassical at late times representing a macroscopic
universe, we chose a Gaussian profile peaked around large
$\omega^\star$: $\tilde{\Psi}_n =
\exp(-(\omega_n-\omega^\star)^2/(2\sigma^2))$.
Since  the contribution from eigenfunctions corresponding to
$\omega$ sufficiently far away from $\omega^\star$ can be neglected, to
 calculate $\Psi(v,\phi)$ via \eqref{eq:psi-int} one has
to only sum a finite number of terms. In numerical simulations the
summation in \eqref{eq:psi-int} was restricted to $n$ such that
$\omega^\star-10\sigma < \omega_n < \omega^\star+10\sigma$.

To calculate the initial data from the WDW coherent state we
evaluated \eqref{sc}. Now the eigenfunctions $e_k(v)$ are
appropriately normalized Bessel functions $\mathcal{K}_{ik}$. They
were calculated using methods of Gil, Segura and Temme \cite{gst}.
As in the initial data construction discussed in the last para, it
is sufficient to restrict the domain of integration to a compact
set $[k^\star-10\sigma; k^\star+10\sigma]$. Resulting integral was
then numerically evaluated using a simple trapezoid method with
set of $10^4$ `probing' points distributed uniformly. On the one
hand, this calculation immediately leads to the solution to the
\WDW equation presented in Fig. \ref{fig:wdw1}. On the other, it
provides us LQC initial data for any lattice $\La_{\vep}$ which
was evolved using (\ref{hc2}).

Eq. \eqref{hc2} constitutes a countable number of ordinary
differential equations. Its domain in $v$ is the lattice
$\La_{\varepsilon}$. However, as noted in section \ref{s5.1}, the
symmetry of the wave function allows us to restrict the
calculations to the sublattice $\La_{+|\vep|}$ for generic $\vep$
and to the part $v>0$ in the exceptional cases $\vep=0$ or $2$.
Due to technical limitations the size of the domain was restricted
by requiring that its elements $v_i$ satisfy the inequality
$|v_i-\vep| \leq 4N\gg v_r(\omega^\star)$. To ensure that the
system remains closed, we impose boundary conditions on the
outermost points $|v_i-\varepsilon| = 4N$ in the generic case and
at the right outermost point $v_i -\vep =4N$ in the exceptional
cases. Since these points lie deep in the classically forbidden
region for all the eigenvalues contributing significantly to the
state, one can safely impose reflective boundary conditions
\begin{equation}
  \Psi(\pm4N+\varepsilon,\phi)\
  =\ \partial_{\phi}\Psi(\pm4N+\varepsilon,\phi)\
  =\ 0 \ .
\end{equation}
In actual simulations we chose  $4N \approx 1.3v_r(\omega^\star)$
which was sufficient for the boundary condition to not affect the
dynamics.

\begin{figure}[tbh!]
  \begin{center}
    \includegraphics[width=4in,angle=0]{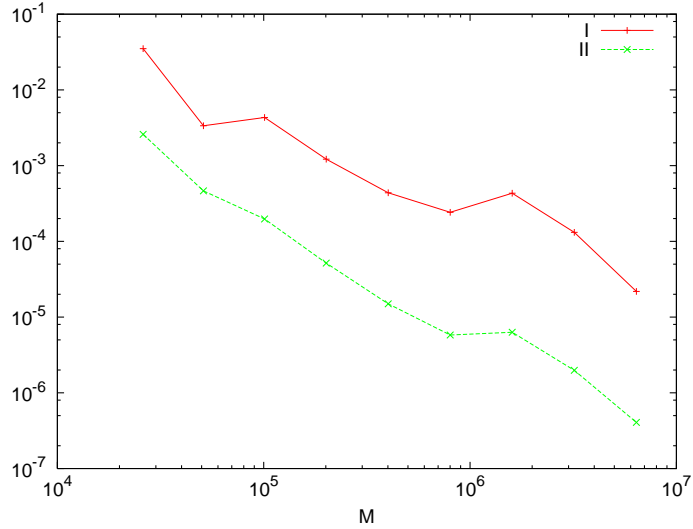}
    \caption{Error functions $\|\Psi_{(M)} - \Psi_{}\|/\|\Psi\|$
      (upper curve) and $\||\Psi_{(M)}| - |\Psi_{}|\|/\|\Psi\|$ (lower
      curve) are plotted as a function of the number of time steps. Here
      $\Psi_{(M)}$ refers to final profile of wave function for a
      simulation with $M$ time steps. $\Psi_{}$ is the limit of final
      profile as $1/M \rightarrow 0$ calculated via polynomial
      extrapolation. In both cases, the evolution began at $\phi=0$ and
      the final profile is evaluated at $\phi =
      -1$. Values of parameters are $p_\phi^\star = 1000$,
      $\Delta p_\phi/p_\phi = 0.018$, $v^\star = 4000$ and
      $\vep = 0$.}
    \label{fig:conv-test}
  \end{center}
\end{figure}

The resulting finite set of equations was integrated using
adaptive $4$th order Runge-Kutta method. To estimate the numerical
errors due to discretization in $\phi$, restrictions
$\Psi|_{\phi}$ calculated for different step sizes were
compared using sup-norm
\begin{equation}
  \|f\|(\phi)\ =\ \sup_{|v_i-\varepsilon|\leq 4N} f(v_i,\phi) \ .
\end{equation}
The step sizes $\Delta\phi$ were chosen to satisfy the inequality
\begin{equation}
  \| \Psi_{\Delta\phi}-\Psi_{\Delta\phi/2} \|
  \leq \,\, \boldsymbol{\epsilon}\,\, \| \Psi_{\Delta\phi/2} \|
  \,\, \Delta\phi \,
\end{equation}
for small preset $\boldsymbol{\epsilon}$, where
$\Psi_{\Delta\phi}$ and $\Psi_{\Delta\phi/2}$ are profiles
calculated with step sizes $\Delta\phi$ and $\Delta\phi/2$
respectively. The dependence of $\|
\Psi_{\Delta\phi}-\Psi_{\Delta\phi/2} \|$ on the number of steps
of the integration for one of the calculations is presented in
Fig. \ref{fig:conv-test}. It shows that the numerical errors
manifest themselves mainly in phases. The differences between the
absolute values of the wave function profiles are approximately
one order of magnitude smaller. Thus the expectation values and
dispersions of observables ${|\hat{v}|_{\phi}}$ are determined
with much better precision than $\Psi$ itself.

\begin{figure}[tbh!]
  \begin{center}
\includegraphics[width=5in,angle=0]{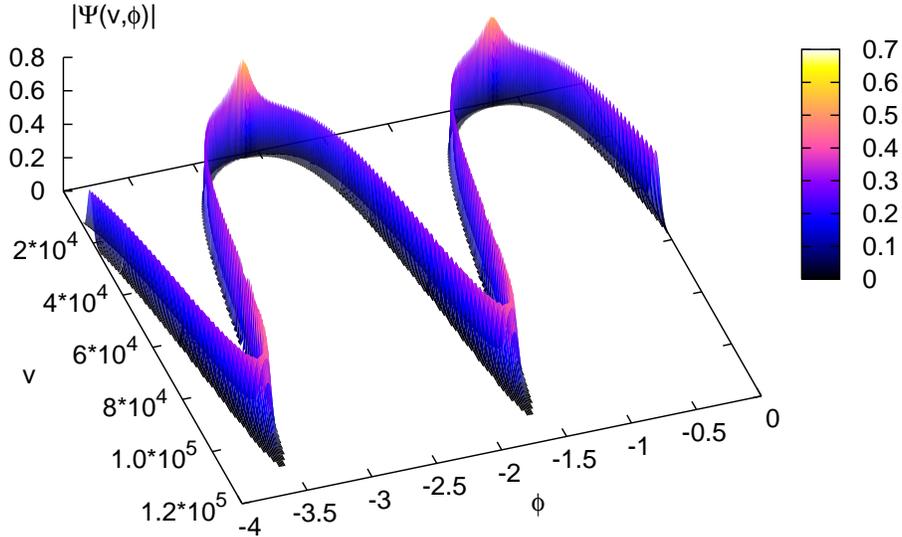}
\caption{The absolute value of the wavefunction obtained by
    numerical evolution. For visual clarity, only values
    $|\Psi| > 10^{-4}$ are shown. Parameters of the initial data
    are: $p_\phi^\star = 5\times 10^3$,\, $\Delta p_\phi/p_\phi =
    0.018,\, v^\star= 5\times 10^4$ and $\vep = 0$. Because of bounces
    the quantum universe exhibits a `cyclic' character.}
    \label{fig:lqc-3d}
\end{center}
\end{figure}

  \begin{figure}[tbh!]
  \begin{center}
    \includegraphics[width=5in,angle=0]{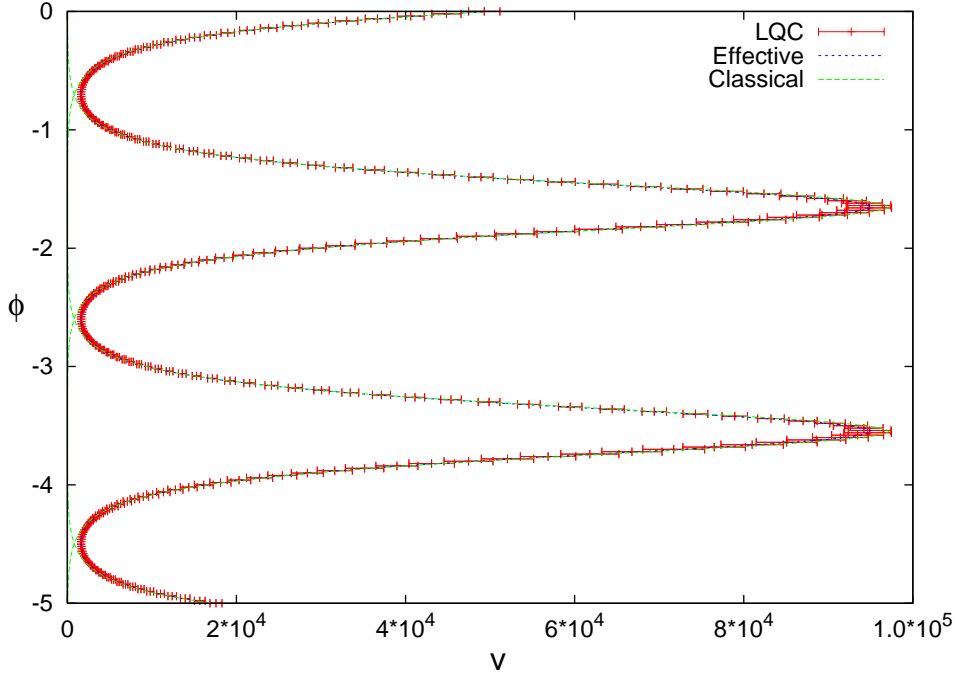}
    \caption{Expectation values and dispersion of
${{|\hat{v}|_\phi}}$ for the wavefunction in Fig. \ref{fig:lqc-3d}
are compared with the classical trajectory and the trajectory from
effective Friedmann dynamics obtained from Eqs.(\ref{dotv}) and
(\ref{dotphi}). The classical trajectory deviates significantly
from the quantum evolution at Planck scale and evolves into
singularities. The effective trajectory provides an excellent
approximation to quantum evolution at all scales.}
\label{fig:traj}
  \end{center}
\end{figure}

The resulting wave functions $\Psi(v,\phi)$ were finally used to
calculate the expectation values $\left<\hat{p}_{\phi}\right>$,
$\left<\right.\!\!{|\hat{v}|_{\phi}}\left.\!\!\right>$ of
observables defined by \eqref{dirac4}. With the inner product
$\left<\Psi|\Psi\right>_{\vep}$ given by \eqref{ip2} they are
equal to the following sums over $\La_{\vep,N} :=
\{v=\pm\vep+4n;-N\leq n\leq N\}$
\begin{subequations}\label{eq:num-expect}\begin{align}
  \left<\Psi\right|{|\hat{v}|_{\phi}}\left|\Psi\right>
    &= \left<\Psi|\Psi\right>_{\vep}^{-1}
    \sum_{v\in\La_{\vep,N}}
    B(v) |v||\Psi(v,\phi)|^2\\
  \left<\Psi\left|\widehat{p_{\phi}}\right|\Psi\right>
    &= \left<\Psi|\Psi\right>_{\vep}^{-1}
    \sum_{v\in\La_{\vep,N}}
    B(v) \bar{\Psi}(v,\phi)(-i\hbar)\partial_{\phi}\Psi(v,\phi)
\end{align}\end{subequations}
The dispersions corresponding to considered observables were
calculated via
\begin{subequations}\label{eq:num-disp}\begin{align}
  \left<\Delta{\hat{p}_{\phi}}\right>^2
    &= \left<\right.\!\!{\hat{p}_{\phi}^2}\!\!\left.\right>
    - \left<{\hat{p}_{\phi}}\right>^2  \\
  \left<\right.\!\!\Delta{|\hat{v}|_{\phi}}\!\!\left.\right>
    &= \left<\right.\!\!{\hat{v}^2_{\phi}}\!\!\left.\right>
    - \left<\right.\!\!{|\hat{v}|_{\phi}}\!\!\left.\right>^2
\end{align}\end{subequations}
where the expectation values
$\left<\right.\!\!{\hat{p}_{\phi}^2}\!\!\left.\right>$ and
$\left<\right.\!\!{\hat{v}^2_{\phi}}\!\!\left.\right>$ are of the
form  \eqref{eq:num-expect}.

In actual simulations $\omega^{\star}$ ranged from $500$ to
$10^4$, with relative uncertainties $\Delta\omega/\omega$ between
$7\times 10^{-3}$ and $2.5\times 10^{-2}$. Wave functions were
calculated on lattices $\La_{\varepsilon}$ corresponding to $5$
different values of $\varepsilon$. Peaks in $v$ of initial
profiles  were chosen to be no smaller than half of the maximal
$v$ predicted by classical theory. For example for
$p_{\phi}^{\star}=500$, $v^{\star}\approx 2000$ whereas for
$p_{\phi}^{\star} = 5000$ the value of $v^{\star}$ ranged between
$5 - 6\times 10^4$. Evolution was performed in both forward and
backward direction.
Representative results of numerical evaluation of $|\Psi(v,\phi)|$
and the expectation values of ${|\hat v|_\phi}$ are shown in
Figs. \ref{fig:lqc-3d} and \ref{fig:traj}. Detailed discussion of the
properties of $|\Psi(v,\phi)|$, the comparison of classical and
quantum evolution and a summary of our results is presented in
Sec. \ref{6.2}.

\section{Physical Implications}
\label{s6}

In this section we discuss the physics of our numerical
simulations. We first obtain the quantum corrected effective
Friedmann equation and show that it is an excellent approximation
to the numerical quantum evolution. We then list the numerical
results and compare and contrast the exact quantum evolution, the
effective theory and the classical Friedmann dynamics.

\subsection{Effective Equations}
\label{s6.1}

The right hand side of the standard Friedmann equation
\be \label{classf} H^2 \equiv \big(\f{\dot{a}}{a})^2  \equiv
\big(\f{\dot{v}}{3v})^2\, =\, \f{8\pi G}{3}\, \rho \, -\,
\f{1}{a^2}\ee
vanishes if and only if  $\rho = 3/8\pi G a^2$. In any classical
solution, at this point the scale factor reaches its maximum
value, the matter density its minimum value, and the classical
dynamics exhibits a turning point from an expanding phase to a
contracting phase. Our numerical simulations show that quantum
dynamics retains this turning point but also gives rise to
additional turning points, resolving the big bang and the big
crunch singularities (see Fig. \ref{fig:lqc-3d}). To gain an
analytical understanding of the physics underlying this
phenomenon, in this subsection we will obtain the leading LQC
corrections to the classical Friedmann equation. This quantum
corrected or `effective' Friedmann equation already suffices to
describe the behavior of the peak of wave functions that we found
numerically in section \ref{s5} (see Fig. \ref{fig:traj}).

The procedure for obtaining these effective equations is the same
as that in the $k$=0 case. Let us begin with a brief summary of
the main ideas. One begins with a geometrical formulation of
quantum mechanics in which the space of pure quantum states is
represented an infinite dimensional symplectic manifold
$\Gamma_{\rm quan}$ ---called the \emph{quantum phase space} (see,
e.g., \cite{as}). The quantum phase space has the structure of a
fiber bundle: the base space is the classical phase space
$\Gamma_{\rm class}$ and the infinite dimensional fiber over any
point $(q_o,\,p_o)$ of $\Gamma_{\rm class}$ is the space of
quantum states in which the expectation values of the canonically
conjugate operators $(\h{q},\,\h{p})$ are $(q_o,p_o)$.%
\footnote{In LQC (and LQG) an important subtlety arises because
there is no operator corresponding to the configuration variables,
i.e., the connections. One has to use holonomies instead. This
issue is handled in \cite{jw}.}
Interestingly, the exact quantum dynamics provides a Hamiltonian
flow on the symplectic manifold $\Gamma_{\rm quan}$ (the
corresponding Hamiltonian being just the expectation value
function of the quantum Hamiltonian operator). To obtain the
desired, first order quantum corrections, one finds a
cross-section of $\Gamma_{\rm quan}$ ---i.e., an embedding of
$\Gamma_{\rm class}$ into $\Gamma_{\rm quan}$--- to which this
flow is approximately tangential in a well-defined sense
\cite{jw}. This approximate quantum Hamiltonian flow unambiguously
projects down to $\Gamma_{\rm class}$ and provides the desired
corrections to classical equations of motion. In a certain sense
this procedure encapsulates the more familiar `effective action'
calculations in the Hamiltonian framework \cite{BS}. In LQC, these
quantum corrections have been obtained for various matter sources
\cite{jw,vt}, where judiciously chosen generalized coherent states
are used to define the required embedding of $\Gamma_{\rm class}$
into $\Gamma_{\rm quan}$. Just as the standard effective action
refers to the in and out vacuum states, the final effective
Hamiltonian (and thus the Friedmann and Raychaudhuri equations)
depend on the specific choice of coherent states. However, the
first order corrections we are interested in are insensitive to
these details. In our case, the resulting effective Hamiltonian
is:
\be\label{heff0} \heff := \f{C_{\rm eff}}{16\pi G} = \f{A(v)}{16
\pi G} \bigg[\sfnsqc - \sfnsq + (1 + \gamma^2) \f{\bar \mu^2 \,
\lo^2}{4} \bigg] + \, (\f{8\pi\g\lp^2}{6})^{-\fs{3}{2}}\, B(v)\,
\f{p_\phi^2}{2} ~. \ee
where $A(v)$ denotes eigenvalues (\ref{A}) of $\hat A$ .

Since the Friedmann equations involve $\dot{a}/a$, to obtain
modifications, we first derive the Hamilton's equations of motion:
\bq\label{dotv} \dot v &=&  \nonumber \{v,\,\heff\} = - \f{8 \pi
\gamma G}{3}\, \f{\partial\heff}{\partial c}\, \, \f{\partial
v}{\partial p}\\
&=& - \f{\gamma \mb A(v)}{2} \,\, \big(\f{8 \pi \gamma
\lp^2}{6}\big)^{-1} \, K^{2/3} \, |v|^{1/3} \, \sfn \cfn \eq
and
\be \label{dotphi} \dot \phi = \{\phi,\,\heff\} = \big(\f{8 \pi
\gamma \lp^2}{6}\big)^{-3/2} \, B(v) \, p_\phi ~. \ee
%

Using the constraint equation $\heff = 0$, i.e.,
\be \sfnsqc = \bigg(\sfnsq - (1 + \gamma^2) \, \f{\mb^2 \lo^2}{4}
\bigg) - 8 \pi G \, \, (\f{8\pi\g\lp^2}{6})^{-\fs{3}{2}}\, B(v)\,
\f{p_\phi^2}{A(v)}\, , \ee
we can eliminate the dependence on the connection $c$ in
(\ref{dotv}) and obtain the desired quantum-corrected Friedmann
equation
\bq \label{quanf} H^2 &=& \nonumber \f{\gamma^2 \mb^2 A(v)^2}{(8
\pi \gamma \lp^2)^2} \, \left(\f{K}{|v|}\right)^{4/3} \,
\Bigg[\chifn - 8
  \pi G \, (\f{8\pi\g\lp^2}{6})^{-\fs{3}{2}}\, B(v) \f{p_\phi^2}{A(v)} \Bigg] \\
&& \, \times \,  \Bigg[1 - \chifn + 8
  \pi G \, (\f{8\pi\g\lp^2}{6})^{-\fs{3}{2}}\, B(v) \f{p_\phi^2}{A(v)}
  \Bigg]\, ,
\eq
which, as with the classical Friedmann equation (\ref{classf}),
involves only $v,\dot{v}$ and $p_\phi$. To take the classical
limit, we note that the volume is given by $(8\pi \g/6)^{3/2}\,
(|v|/K)\, \lp^3$ and that (the area gap and hence) $\mb$ goes to
zero in this limit. Therefore, in the limit (\ref{quanf}) reduces
precisely to (\ref{classf}). Terms containing $\mb$ represent the
quantum geometry corrections.

In view of the fact that the quantum equations are invariant under
the orientation reversal map $\Pi\, \Psi(v) = \Psi(-v)$, it will
suffice to restrict ourselves to $v\ge 0$. Now, for $v>1$ we have
\be A(v) = - \f{\sqrt{48 \pi}}{\gamma^{3/2} \mb^2} \,
\left(\f{|v|}{K}\right)^{1/3} \, \lp \ee
and for $v\gg 1$, we have:
\be B(v) =
\f{K}{v} + O\big(v^{-3}\big)
\ee
%
Therefore, the leading order quantum corrected equation is given
by:
\bq\label{modfeq1} H^2 &=& \nonumber  \left(\f{8 \pi G}{3} \rho +
\f{1}{\gamma^2 \mb^2} \bigg(\f{8 \pi \gamma \lp^2}{6} \bigg)^{-1}
\, \bigg(\f{K}{|v|}\bigg)^{2/3} \,\bigg(\chifn\right)\bigg) \\
&&\, \times \, \left(1 - \chifn - \f{\rho}{\rcr} \right) +
O\big(v^{-3}\big) \eq
where $\rho$ denotes the eigenvalues of the energy density
operator $\hat \rho \equiv \widehat{p_\phi^2/|p|^3}$ and as in the
$k$=0 case we have set $\rcr = 3/(16 \pi^2 \gamma^3 G^2 \hbar) \,
\approx \, 0.82 \rho_{\mathrm{Pl}}$. Although for brevity we have
kept the sine functions in this equation, since $v^{-1} \sim
\mb^{3}$, to the leading order considered here one only needs to
keep terms $O(\mb^8)$ in their Taylor expansions. Finally, note
that on substituting $\lo = 0$ in (\ref{modfeq1}), we immediately
obtain the effective Friedmann equation of \cite{aps3} for the
$k$=0 model.

To probe the possible turning points, it is useful to rewrite the
modified Friedmann equation (\ref{modfeq1}) in the following form
\be H^2 = \f{8 \pi G}{3} \, (\rho - \rho_1) \,
\left(\f{1}{\rcr}(\rho_2 - \rho)\right) + O(v^{-3})\ee
with,
\be \rho_1(v) = - \f{3}{8 \pi G} \f{1}{\gamma^2 \mb^2} \,
\left(\f{8 \pi \gamma \lp^2}{6} \right)^{-1} \,
\left(\f{K}{|v|}\right)^{2/3} \, \left(\chifn\right) \ee
and
\be \rho_2(v) = \rcr \left(1 - \chifn\right) ~. \ee
Note that while $\rcr$ is a constant, $\rho_1, \rho_2$ are
functions of $v$ (and $\rho$ is a function of $v,\, p_\phi)$.
Along each dynamical solution of the effective evolution equations
(\ref{dotv}) and (\ref{dotphi}), $\rho, \rho_1,\rho_2$ all evolve.
In the effective dynamics, along any given dynamical trajectory
turning points occur when $\rho=\rho_1$ or $\rho=\rho_2$. Plots of
solutions to the effective equations show that the classical
recollapse occurs when $\rho=\rho_1$, where the universe reaches
its maximum radius $a_{\rm max}$ and minimum density $\rho_{\rm
min}$, and the quantum bounce occurs when $\rho=\rho_2$ where the
universe reaches its minimum radius $a_{\rm min}$ and maximum
density $\rho_{\rm max}$. Now, an examination of the expressions
of $\rho_1$ and $\rho_2$ show
\be \rho_{\rm min} := \rho_1|_{a=a_{\rm max}}= \f{3}{8\pi G a_{\rm
max}^2}\,\left(1 + O\left(\f{\lp^4}{a^4_{\rm max}}\right)\right)\quad {\rm and}
\quad \rho_{\rm max} := \rho_2|_{a=a_{\rm min}} =  \rcr\, \left(1+
O\left(\f{\lp^2}{a_{\rm min}^2}\right)\right) ~. \ee
Numerical simulations showed that our notion of
`semi-classicality at late times' is surprisingly weak. For
example, in the simulation with $p_\phi \approx 5\times 10^3\hbar$
(in the classical units $c$=$G$=1), the universe grows only to a
maximum radius of $\approx 23 \lp$ before undergoing the classical
recollapse. Even for this small universe, effective equations
predict that the density $\rho_{\rm min}$ at the recollapse should
agree with the classical Friedmann formula $\rmin = {3}/{8\pi G a_{\rm
max}^2}$ to one part in $10^{-5}$ and the density $\rho_{\rm max}$
at the quantum bounce would equal the critical density $\rcr$
---the density at the bounce in the $k$=0 models--- to within a
couple of percent. \emph{These predictions are borne out in the
numerical simulations of the exact LQC equations.} Since $a_{\rm
max}$ scales as $(p_\phi)^{1/2}$ and $a_{\rm min}$ as
$(p_\phi)^{1/3}$, effective equations imply that both
approximations improve as $p_\phi$ increases and become almost
exact for universes which grow to interesting macroscopic sizes.

In particular, as one would hope, the effective theory accurately
reproduces the predictions of classical general relativity in the
low curvature regime. Yet, already the leading order correction
from quantum geometry is strong enough to resolve singularities
and replace them with a bounce. It is also noteworthy that for
universes which grow to large macroscopic sizes, the density at
the quantum bounce is universal, and equals that in the $k$=0
models. These general features represent key predictions of the
numerical evolution of the full quantum equations. The effective
theory provides a physical understanding of how they come about.
In particular it shows that the most important corrections come
from the gravitational part of the quantum Hamiltonian constraint.
Indeed, since we need only the leading, classical part of $B(v)$
in this analysis, the quantum modifications of the matter
Hamiltonian play no role in the leading order corrections
discussed here.

In the last part of our above discussion we have ignored terms
$O(v^{-3})$. A priori it is possible that the sum of these terms
is not negligible and could even dominate the leading term.
However, a comparison with numerical simulations shows that the
predictions obtained from just the leading order quantum
corrections in the effective theory accurately describe the full
quantum dynamics of states which are semi-classical at late times
and the accuracy improves as $p_\phi$ increases. Thus, although
one could not draw definitive conclusions just from effective
equations, when used in conjunction with numerical simulations,
they provide an easily manageable and powerful tool to probe
quantum geometry effects for universes which grow to $a_{\rm max}
\sim 25 \lp$ or more.

\begin{figure}[]
  \begin{center}
    $a)$\hspace{8cm}$b)$
    \includegraphics[width=3.2in,angle=0]{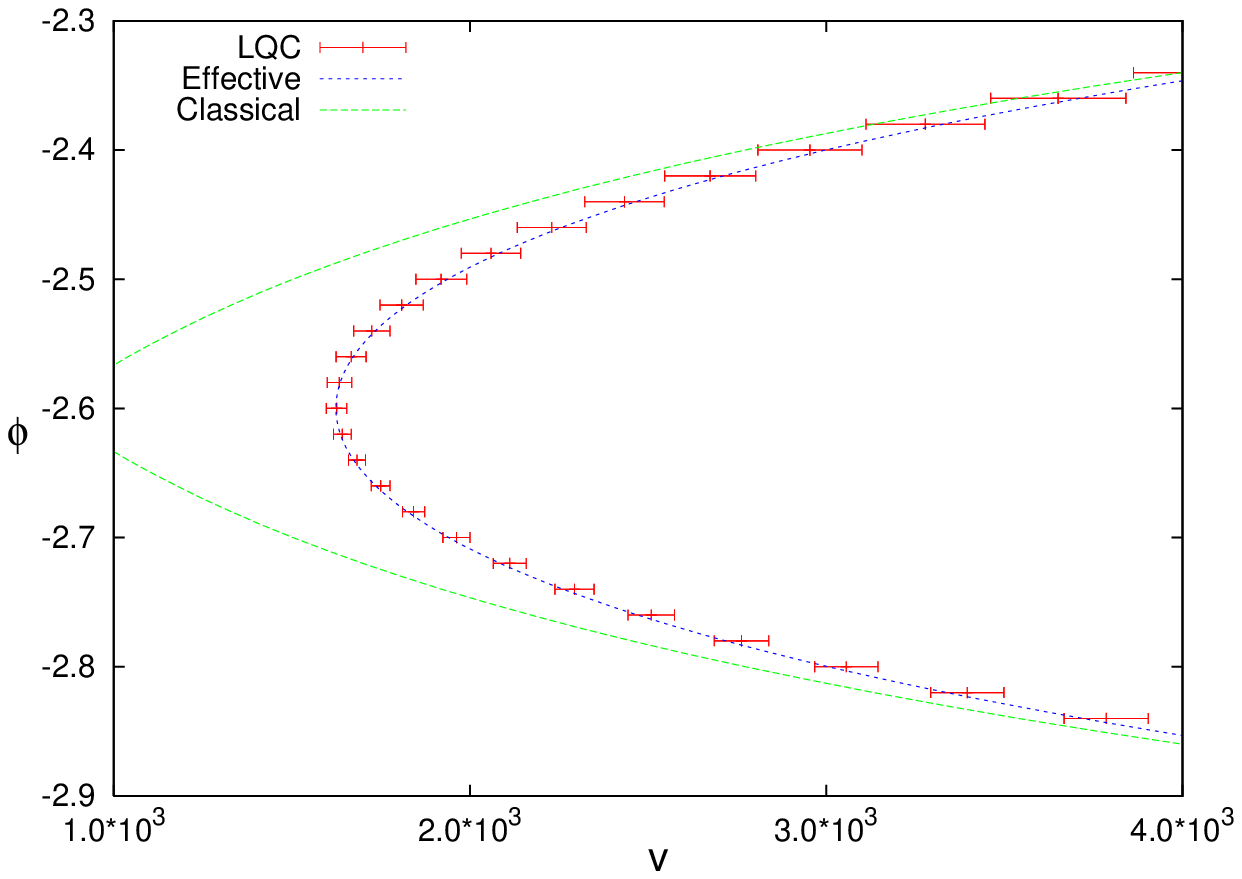}
    \includegraphics[width=3.2in,angle=0]{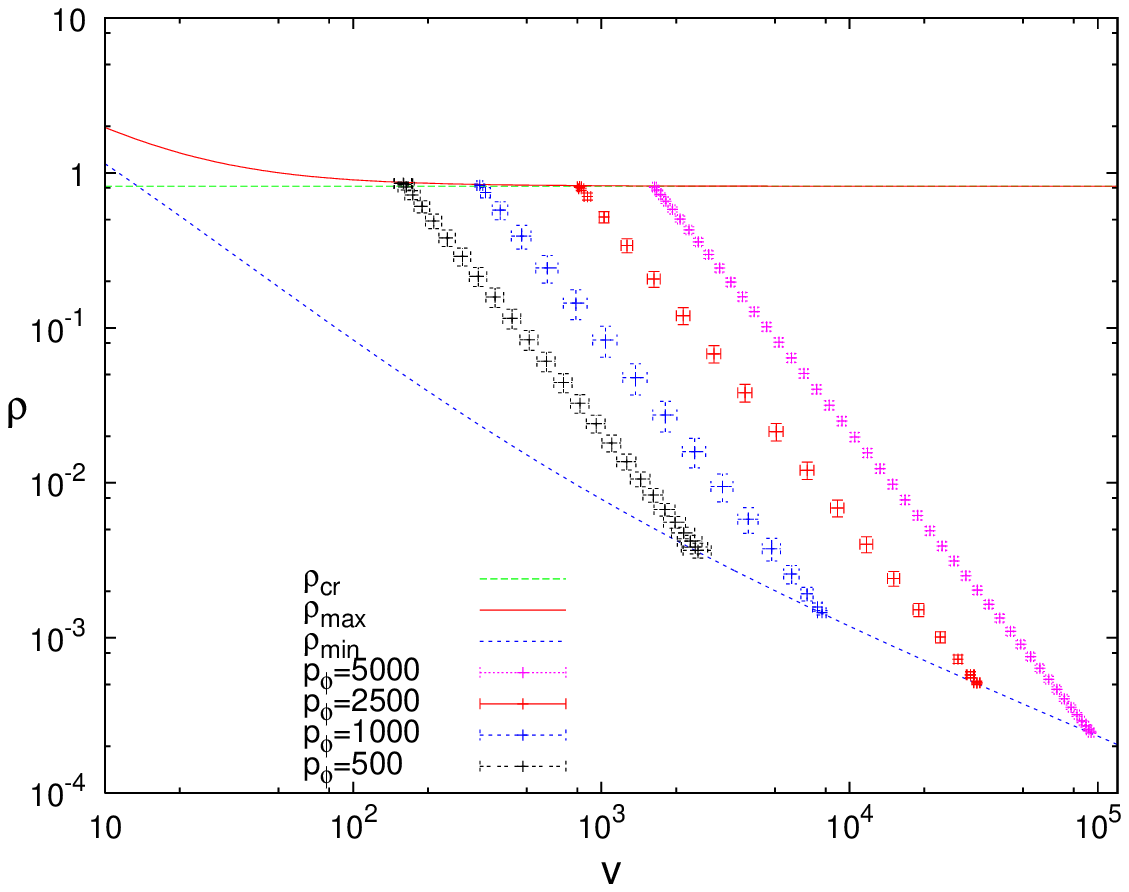}
\caption{$a)$ Zoom on the portion near the bounce point of
comparison between the expectation values and dispersion of
${\hat{v}|_{\phi}}$, the classical trajectory and the solution to
effective dynamics. At large values of $|v|_\phi$ the classical
trajectory approaches the quantum evolution. Values of parameters
are the same as in Fig. \ref{fig:traj}.\,\, $b)$ The behavior of
expectation values of $\hat \rho$ for different values of
$p_\phi^\star$  are shown. On each solution, these are bounded
between a $\rmax$ and a $\rmin$. For a universe peaked at large
values of $p_\phi$, $\rmax \approx \rcr$.} \label{fig:density}
  \end{center}
\end{figure}

\subsection{Results}
\label{6.2}

Main results on quantum dynamics can be summarized as follows.

\begin{enumerate}

\item Consider a classical solution which evolves from the
big-bang to the big crunch, reaching a large maximum radius
$a_{\rm max}$. Fix a point on this trajectory  where the universe
has reached macroscopic size and consider a semi-classical state
peaked at this point (see section \ref{s5.2}). Such states remain
sharply peaked throughout the given `cycle', i.e., from the
quantum bounce near the classical big-bang to the quantum bounce
near the classical big-crunch. The notion of semi-classicality
used here is rather weak: these results hold even for universes
with $a_{\rm max} \approx 25 \lp$ and the `sharply peaked'
property improves as $a_{\mathrm{max}}$ grows.

\item The trajectory defined by the expectation values of the
Dirac observable $\h{v}|_\phi$ in the full quantum theory is in
good agreement with the trajectory defined by the classical
Friedmann dynamics until the energy density of the scalar field
becomes comparable to the maximum energy density $\rmax \sim \rcr
\approx 0.82\rho_{\rm Pl}$. Then the classical trajectory deviates
from the quantum evolution. In the classical solution, the matter
energy density keeps increasing on further evolution, eventually
leading to a big bang (respectively, big crunch) singularity in
the backward (respectively, forward) evolution, when $v
\rightarrow 0$. The situation is very different with quantum
evolution. Now the universe bounces at $\rho =\rmax$, avoiding the
past (or the big bang) and future (or the big crunch)
singularities.

\item The expectation values and relative dispersions of
${\widehat{p_\phi}}$ remain constant during different stages of
evolution. Thus the expanding and contracting branches correspond
to the same value of $\langle {\widehat{p_\phi}} \rangle$.
Further, as a check on numerics we verified that the norm of the
states is also preserved during the entire evolution.

\item After the quantum bounce the energy density of the universe
decreases and, when $\rho \ll \rmax$, the quantum evolution is
well-approximated by the classical trajectory. On subsequent
evolution, the universe recollapses both in classical and quantum
theory at the value $v=v_{\mathrm{max}}$ when energy density
reaches a minimum value $\rmin$.

\item The trajectory obtained from effective Friedmann dynamics
(\ref{modfeq1}) is in excellent agreement with quantum dynamics
\emph{throughout the evolution.} (See Figs. \ref{fig:traj} and
\ref{fig:density}a.) In particular, the maximum and minimum energy
densities predicted by the effective description agree with the
corresponding expectation values of the density operator $\hat
\rho \equiv \widehat{p_\phi^2/|p|^3}$ computed numerically.
Evolution of the expectation values of $\hat \rho$ with $|v|_\phi$
is shown in Fig. \ref{fig:density}b.

\item For quantum states under discussion, the density $\rmax$ is
well approximated by $\rcr \approx 0.82 \rho_{\mathrm{Pl}}$ up to
terms $O(\lp^2/a_{\rm min}^2)$, independently of the details of
the state and values of $p_\phi$. (For a universe with maximum
radius of a megaparsec, $\lp^2/a_{\rm min}^2 \approx 10^{-76}$.)
The density $\rmin$ at the recollapse point also agrees with the
value $(3/(8 \pi G a^2_{\rm max})$ predicted by the classical
evolution to terms of the order $O(\lp^4/a_{\rm max}^4)$.
Furthermore the scale factor $a_{\rm max}$ at which recollapse
occurs in the quantum theory agrees to a very good precision with
the one predicted by the classical dynamics.

\item The relative dispersion of $|v|_\phi$ increases ---but very
slowly ---as one evolves through many cycles. Effective Friedmann dynamics
provides insight into this behavior of the quantum state. Let us
first consider two nearby solutions to the effective equation with
slightly different $p_\phi$ but with same value of $v$ at a chosen
$\phi = \phi_o$. Then the relative difference between $v$ of the
two solutions after one cycle can be estimated using
Eqs.(\ref{dotv}) and (\ref{dotphi}) as
\be \label{dvv-est} \f{\delta v}{v} \approx \sqrt{48 \pi^3 G} \,\,
\alpha \,\, \f{\delta p_\phi}{p_\phi} \ee
where $\delta p_\phi$ is the difference between values of $p_\phi$
of the two effective trajectories and $\sqrt{G}\, \alpha\ =\
0.0259272\pm (5 \times 10^{-7})$ (see \ref{eq:dw-est1}). This
estimate was found to provide a good upper bound on the relative
dispersions computed using numerical evolution of the quantum
state.

\item The state remains sharply peaked for a \emph{very large
number of `cycles'.} This number can be estimated using
Eq.(\ref{dvv-est}). Consider the example of a semi-classical state
with an almost equal relative dispersion in $p_\phi$ and $|v|_\phi$
and peaked at a large classical universe of the size of a
megaparsec. When evolved, it remains sharply peaked with relative
dispersion in $|v|_\phi$ of the order of $10^{-6}$ \emph{even
after $10^{50}$ cycles of contraction and expansion!}  Any given
quantum state eventually ceases to be sharply peaked in $|v|_\phi$
(although it continues to be sharply peaked in $p_\phi$).
Nonetheless, the quantum evolution continues to be deterministic
and well-defined for infinite cycles, i.e., on the entire real
line of the emergent time $\phi$. This is in sharp contrast with
the classical theory where the equations break down at
singularities and there is no deterministic evolution from one
cycle to the next. In this sense, in LQC  the $k$=1 universe is
{\it cyclic}, devoid of singularities. As in the $k$=0 case, this
non-singular evolution holds for all states, not just the ones
which are semi-classical at late times. There is no fine tuning of
initial conditions. Also, there is no violation of energy
conditions. Indeed, as discussed in section \ref{s6.1}, quantum
corrections to the matter Hamiltonian do not play any role in the
resolution of the singularity. The standard singularity theorems
are evaded because the geometrical side of the classical
Einstein's equations is modified by the quantum geometry
corrections of LQC.

\item In the $k$=1 model, certain effective equations have been
written down and used to predict non-singular bounces
\cite{st_closed} in the broad framework of LQC. The presence of
these bounces was also used to study the onset of a successful
period of inflation in closed models \cite{qmc_closed}. How do
these analysis compare with that presented in this paper? There
are two important differences. First, these works focused on the
matter part of the Hamiltonian constraint and made a crucial use
of quantum corrections to the matter Hamiltonian arising from the
use of representations (of $\SU(2)$-holonomies) labeled by large
values of $j$. Second, these large $j$ representations were used
only in the matter part of the Hamiltonian and not in the
gravitational part. Since then Perez \cite{ap} has shown that
mathematical consistency requires us to use the $j=1/2$
representation in 3-dimensional gravity. He also argues that the
same should hold in 4 dimensions. In any case, while the use of,
say $j=1$ representation could be justified because the model has
no spinor fields, the use of large $j$ values appears to be
unnatural and needs an independent justification which is still
lacking. Similarly, without an independent justification, an
asymmetric treatment of the gravitational and matter appears to be
ad-hoc \cite{kv}, somewhat similar to using two different metrics on the
right and left hand side of classical Einstein's equations. Our
analysis used the fundamental, $j=1/2$ representation for geometry
as well as matter. Our numerical simulations as well as effective
equations show that with this choice the modifications of the
matter Hamiltonian play no role. This is consistent with findings
in the older literature; indeed this is the reason why large $j$
values were used there. In our analysis, it is the modification of
the gravitational part of the Hamiltonian constraint that plays a
key role in the singularity resolution. In the older literature,
by contrast, these modifications were ignored. Nonetheless, basic
physical ideas in these older works are intriguing and it would be
interesting to reanalyze those issues using the Hamiltonian
constraint introduced in this paper.

\end{enumerate}

We will conclude this section by clarifying two issues that have
arisen from a recent work \cite{pl} addressed to computational
physicists, particularly numerical relativists. Although this
discussion refers only to the $k$=0 model considered in
\cite{aps1,aps2,aps3}, it is included here because the same issues
can arise in the $k$=1 model.

The phrase ``These bounces can be understood as spurious
reflections'' in the abstract (and again in the body) of \cite{pl}
was interpreted by some as suggesting that the bounces reported in
\cite{aps1,aps2,aps3} were artifacts of bad numerics. This is
certainly \emph{not} the case: Not only were those simulations
performed with all the due care but our result that the LQC
equations predict a genuine, physical bounce was in fact
reproduced in the first half of \cite{pl}. Indeed, this part is a
nice summary of our numerical results geared to computational
physicists. From discussions with the author we understand that
the intent of that phrase was to say: `had the physical problem
been to solve a wave equation in the \emph{continuum} and had one
used non-uniform grids, one would also have found bounces which,
from the perspective of continuum physics of this hypothetical
problem, would be interpreted as spurious reflections in finite
difference discretizations'. This is likely to be an illuminating
point for computational physicists but is not physically relevant
in LQC where the basic equation is a difference equation.

The second part of the paper considers a modification of the
quantum Hamiltonian constraint by ``adding ad-hoc higher order
terms.'' It is then suggested that such modifications could remove
the bounce. Let us analyze the issue from a mathematical
perspective even though the analysis and conclusions have no
obvious physical significance since the modifications do not
result from any systematic, physical considerations. Then, since
the physical state is symmetric under orientation reversal,
simulations reported in \cite{pl} imply that the bounce would
\emph{not} disappear but change its character. In the
\emph{physical} solution there would again be a pre-big-bang,
contracting branch which would be joined in a deterministic
fashion to a post-big-bang expanding branch. However, now the two
branches will meet at $v=0$. Although the ensuing differences are
not trivial, the qualitative picture is not changed even by these
ad-hoc
modifications.%
\footnote{Indeed, as the numerical simulations of \cite{aps2,aps3}
show, the `reflection' and transmission' phenomenon discussed in
the second part of \cite{pl} occurs even in standard LQC. It does
not have deep physical significance in the final picture because
physical states are symmetric under the orientation reversal map
$\Pi$.}

\section{Discussion}
\label{s7}

\subsection{Key Features}
\label{s7.1}

Key features of the $k$=1 model can be summarized as
follows.\medskip

i) The scalar field $\phi$ serves as emergent time at all three
levels: classical general relativity, \WDW theory and LQC. In the
classical theory, every solution undergoes a recollapse but $\phi$
remains single valued. Each solution begins with a big bang and ends with a big crunch and $\phi$
ranges over the entire real line irrespective of the constant of motion
$p_\phi$. In the \WDW theory and LQC the form
of the Hamiltonian constraint operator implies that $\phi$ can
serve as emergent time also in the quantum theory. Situation with
the range of $\phi$ in the \WDW theory is the same as that in 
classical general relativity since the singularities are not
resolved, i.e., since quantum dynamics can not  unambiguously
evolve the state across these singularities. In LQC on the other
hand the singularities \emph{are} resolved and the quantum
evolution across the putative classical singularities is
deterministic. The range of $\phi$ continues to be the entire real line.

ii) In LQC, three sets of results show that the big bang and the
big-crunch singularities are both resolved. First, the LQC
effective equations do not break down. Rather, while the classical
Friedmann equation has only one root at which $H^2 =
(\dot{a}/a)^2$ vanishes, quantum corrections introduce a second
root when the matter density enters the Planck regime, altering
classical dynamics and giving rise to bounces. The second result
refers to full quantum dynamics: in contrast to the \WDW theory,
in LQC \emph{every} state in $\Hp$ has a well-defined unitary
evolution for the full range $(-\infty,\, \infty)$ of the
`emergent time' $\phi$. The third set of results is more detailed.
It again involves the full dynamics of LQC but only semi-classical
states in $\Hp$. Consider a classical trajectory in the $v$-$\phi$
plane in which the universe evolves to a macroscopic maximum size.
The classical Friedmann equation implies that the universe attains
its maximum volume $V_{\rm max}$ at the recollapse point and this
value is related to the constant of motion $p_\phi$ via
\footnote{Here and in what follows, numerical values are given in
the classical units G=c=1. In these units $p_\phi$ has the same
physical dimensions as $\hbar$ and the numerical value of $\hbar$
is $2.5\times 10^{-66}{\cm}^2. $}
\be\label{Vmax}  V_{\rm max} = (16\pi G/3 \lo^2)^{\fs{3}{4}}\,
p_\phi^{\fs{3}{2}}\,\, \approx \,\,  0.6\, p_\phi^{\fs{3}{2}} .\ee
Hence, in the solution under consideration $p_\phi$ has to be
large. Consider a point on this classical trajectory at a late
time when the volume is macroscopic and a semi-classical state in
which Dirac observables $\hat p_\phi$ and $|\hat v|_{\phi_o}$ are peaked at
this point at `time' $\phi=\phi_o$. When evolved, this quantum
state remains semi-classical, sharply peaked at the classical
solution under consideration during the entire cycle, except near
the big-bang and the big-crunch. There, because of repulsive
effects of quantum geometry, the wave function bounces when the
peak reaches a minimum volume:
\be V_{\rm min} = \big( \f{4\pi G\g^2 \Delta}{3}
\big)^{\f{1}{2}}\, p_\phi \,\, \approx (1.28\times 10^{-33}\,
{\cm})\,\, p_\phi\ee
Thus, the wave function does not enter a neighborhood of the
classical singularity. The size of this neighborhood is dictated
by the value of the constant of motion $p_\phi$ and can be
\emph{much} larger than the Planck size. Finally, note that in the
mathematical limit in which the area gap $\Delta$ is taken to
zero, $V_{\rm min}$ vanishes. More generally, this is also the
limit in which LQC reduces to the \WDW theory. Thus, one can use
$\Delta$ as a knob to turn on or off quantum geometry effects and
understand the `mechanism' behind the singularity resolution.

iii) To better understand the physical conditions at the quantum
bounce, let us consider a couple of examples. Consider first a
quantum state describing a universe which attains a maximum radius
of a megaparsec. Then the quantum bounce occurs when the volume
reaches the value $V_{\rm min} \approx 5.7 \times 10^{16}\,
{\cm}^3$, \emph{some $10^{115}$ times the Planck volume.} As a
second illustration, consider a semi-classical state representing
a large universe whose density at the classical re-collapse is
about the current density of our universe $\rho \approx 9.7 \times
10^{-30}\, {\rm gm/cc}$. Then, the quantum bounce would occur at
$V_{\rm min} \approx 1.4\, \times 10^{24}\, \cm^3$. Thus the
`quantum' or the `Planck' regime is defined not by the volume of the
universe but by the value of the matter density, or space-time
curvature. In universes which grow to large macroscopic sizes,
these quantities can attain Planck scale even when the volume of
the universe is large. Figures quoted above were arrived at using
our model where the scalar field is massless. The presence of
potentials could significantly modify their values. Still, since
the values of $V_{\rm min}$ are so huge, these considerations are
useful in drawing qualitative conclusions. For example, they
suggest that in universes which grow to macroscopic sizes, the so
called `$d_j$-effects' associated with modifications of the matter
Hamiltonians due to quantum geometry will not be dynamically
significant in homogeneous models (unless one considers
astronomically large ---and hence implausible--- values of $j$).

iv) Since detailed predictions were obtained only for states which
are semi-classical at late times, it is interesting to ask
\emph{how} semi-classical these states have to be. How quantum
mechanical can we make the parameters of the universe, still
keeping the quantum state semi-classical in the sense used in this
paper? The typical values of $p_\phi$ used in the simulations
reported in section \ref{s5} was $5 \times 10^3 \lp^2$. These
universes evolve only to a maximum volume of $V_{\rm max} \approx
2.3 \times 10^5 \lp^3$ before undergoing a recollapse.
\emph{Results of our numerical simulations show that the necessary
semi-classical considerations hold already for such small
universes.} In particular, the maximum value $\rho_{\rm max}$ of
the matter density is well approximated by the critical value
$\rcr = 0.82\, \rho_{\rm Pl}$ already for these universes. A
combination of numerics and effective-equations shows that the
approximation becomes increasingly better as one considers larger
and larger universes. Similarly, our numerical results have shown
that the recollapse occurs at the classically predicted values
already for these universes. In these cases, the matter density
even at the recollapse point is quite high, approximately $2.2
\times 10^{-4}\, \rho_{\rm Pl}$. Thus, several interesting
phenomena occur in a rather small interval (of just four to
five orders of magnitude) of density and volume. As our detailed plots
show, there is a very narrow range of these parameters in which
quantum geometry effects become significant. They grow extremely
quickly, overwhelm the classical attractive force, cause the
bounce and then become insignificant very quickly again.

v) We would like to emphasize that while quantum geometry effects
resolve classical singularities, we did not predict the emergence
of classicality at late times within any given cycle.%
\footnote{ As discussed in the last section of \cite{aps2}, the
bounce picture has a suggestive, intriguing relation to the
Hartle-Hawking \cite{hh} proposal for the wave function of the
universe. It may well suggest an avenue to address this issue.}
These two issues are logically  distinct. Indeed, in our detailed
analysis we simply restricted ourselves to a single cycle and to
states which are semi-classical at late times therein. At first,
it may appear that the problem of actually specifying such states
would be impossibly difficult in more realistic models since such
specification would have to incorporate all the complexities that
have developed during the epoch during which the universe grew to
a macroscopic size. However, as discussed above, this `macroscopic
size' can be very small and the required specification of a
semi-classical state could be done at a relatively early time
before complicated structures develop. For example, in an
inflationary scenario one could specify the state immediately
after the end of inflation, or perhaps even before the onset of
inflation since, in the current observationally favored scenarios,
matter density is significantly smaller than the Planck density
even at the onset. Thus, while the conceptual issue of singling
out a preferred family of states using general principles remains
largely unexplored, there do not appear to be any `practical'
difficulties in specifying the state.

vi) Since both the big bang and the big crunch singularities are
resolved, the quantum wave function evolves through
\emph{infinitely many} classical cycles. Thus, in the $k$=1 model,
the quantum space-time of LQG is vastly larger than that the
classical space-time of general relativity. The issue of emergence
of semi-classicality raised above can now be elevated to the
\emph{infinite} history of the quantum universe. Let us then begin
a classical solution in which the maximum volume of the universe is
large, say more than a megaparsec, and consider a quantum state which
is peaked at a point on this classical trajectory at a late `time',
$\phi=\phi_o$. As discussed in section \ref{s6}, this state has
interesting properties. Except near the classical singularities, it
will remain sharply peaked not only on the given classical solution
during its cycle but also for over $10^{50}$ cycles resulting from
quantum bounces. However, because the eigenvalues of $-i\p_\phi \equiv
\sqrt{\Theta}$ are not exactly evenly spaced, eventually the wave
function will spread and cease to be semi-classical. The
issue of converse is intriguing.  Consider any state which is sharply
peaked at a large value of the constant of motion $\hat{p}_\phi$ but
has a large spread for the volume operator $|\hat{v}|_\phi$ at the
initial instant of `time' $\phi=\phi_o$.  Initially, such a state is
not semi-classical. However, would the LQC dynamics evolve it to a
state which is eventually sharply peaked at a classical trajectory?
The answer appears to be in the affirmative \cite{tp}. If so, all
states which are peaked at a very large value of $p_\phi$ would
eventually become semi-classical. In this precise sense,
semi-classicality would be generic.

\subsection{Classical recollapse from LQC}
\label{s7.2}

As discussed in section \ref{s1}, a major challenge to any
background independent quantum gravity approach, such as LQG, is
to ensure that there is a sufficiently large semi-classical
sector. LQC offers a non-trivial context to probe this issue. Are
the quantum geometry effects subtle enough to dominate near
classical singularities but turn themselves off on large scales?
As results of \cite{aps2} show, this is a delicate issue. Indeed,
the evolution generated by the Hamiltonian constraint that was
generally used in LQC until recently (the so-called
`$\mu_o$-evolution'), the answer was in the negative. Certain
quantum effects associated with that evolution could alter the
classical predictions even in regimes in which the matter
densities and space-time curvatures are completely tame. In the
$k$=0 case, this severe drawback was overcome by the `improved
dynamics' of \cite{aps3}.

In the $k$=1 models, the classical recollapse provides an
excellent venue to test semi-classical viability because, as the
analysis of section \ref{s6} shows, both the classical recollapse
and the quantum bounce are governed by the same condition:
the vanishing of $H^2 = (\dot{a}/a)^2$. For the recollapse, an agreement
with the classical theory requires that quantum geometry effects
be negligible while the bounce can occur only if these effects
dominate. At first sight then there appears to be a tension. In
\cite{gu}, Green and Unruh analyzed this issue numerically in the
same model as the one considered in this paper and concluded that
the tension is real. More precisely, they used \emph{the then
available} Hamiltonian constraint of LQC \cite{closed} and found
numerical evidence against the occurrence of recollapse. For large
universes, the classical recollapse occurs when matter density and
space-time curvatures are very small compared to the Planck scale.
Therefore, a theory in which recollapse does not occur would
contradict classical general relativity in a domain where there is
every reason to expect its validity. Although Green and Unruh did
not have access to a physical inner product or observables to
arrive at a definitive interpretation of their results, they
concluded that it is unlikely that one could find an
interpretation in which such large deviations from the classical
theory are appropriate.

In this paper we have overcome important limitations pointed out
by Green and Unruh, thereby completing the LQC program. We found
that LQC does predict a recollapse and, furthermore, it occurs at
the values of matter density and volume predicted by the classical
theory. How did this strikingly different conclusion come about?
We will conclude by first summarizing how the general criticisms
of \cite{gu} were addressed and then discussing the issue of the
recollapse.\medskip

i) General Framework: Green and Unruh began by pointing out that
much of the then successes of LQC arose from effective equations
and the interpretation of quantum states had remained unclear
because the physical inner product and observables had not been
specified. They pointed out that, in particular, the issue of time
had not been addressed explicitly and the implicit use of the
scale factor as time has obvious problems in the $k$=1 model. In
this paper we showed that the use of the scalar field as emergent
time is free of the difficulties associated with the multi-valued
character of the scale factor in closed models. We constructed the
physical sector of LQC in detail, including the physical inner
product, Dirac observables and well-controlled semi-classical
states. We then numerically solved the Hamiltonian constraint and
calculated the expectation values and fluctuations of Dirac
observables. In particular, we analyzed dynamics in full LQC, not
in just in an effective approximation.

ii) Recollapse: Our Hamiltonian constraint is quite different from
that used in \cite{gu}. Thus, the very starting points of the two
sets of numerical simulations are distinct. In particular, while
$\h{C}_{\rm grav}$ used in \cite{gu} was \emph{not self-adjoint},
ours is. This has two consequences. First, since our operator
$\Theta$ is self-adjoint on $\Hkg$ and we are guaranteed that it
admits a complete set of eigenfunctions. Furthermore since its
spectrum is discrete \cite{warsaw} the eigenfunctions in the
spectral family are normalizable. This immediately implies that
they must decay for large $|v|$. Our numerical simulations showed
that the decay is exponential for large $|v|$, which ensured that
physically appropriate wave packets would exhibit a recollapse.
The numerical task of actually finding these eigenfunctions was
delicate because the normalizable ones constitute a set of `zero
measure' among all eigenfunctions, i.e., because `most'
eigenfunctions diverge in at least one of the two asymptotic
regimes $v \rightarrow \pm \infty$. However, because $\Theta$ is
self-adjoint with appropriate properties, the existence of these
eigenfunctions was ensured from the beginning. This was not the
case for the analysis of \cite{gu}. Indeed the principal argument
there was that the authors found only exponentially growing
eigenfunctions, signaling absence of recollapse. Secondly, for
reasons discussed in Sec.\ref{s1}, it was important that we used the
`improved dynamics' of \cite{aps3} (rather than the older,
`$\mu_o$ evolution'). Finally, the effective-equation-analysis
shows how LQC manages to have both the quantum bounce and the
classical recollapse. For, the modified Friedmann equation now has
two roots: one a la classical general relativity at a low density
causing the recollapse, and a \emph{new one} near Planck density
causing the bounce. The quantum geometry effects are small at the
old, classical root but cause and thus dominate the new root.

\bigskip
\textbf{Acknowledgments:} We would like to thank Golam Hossain and
Pablo Laguna for discussions, Eloisa Bentivegna for her help in
calculating the Bessel functions of section \ref{s3} and
participants of the 11th Marcel Grossmann conference and the
Unruh-Wald fest for interesting comments and suggestions. This
work was supported in part by the NSF grant PHY-0456913, the
Alexander von Humboldt Foundation, the Kramers Chair program of
the University of Utrecht, the Eberly research funds of Penn State
and the Marie Curie Incoming International Fellowship
MIF1-CT-2006-022239 to KV. 

\begin{appendix}

\section{Invariant frames, explicit charts and holonomy}
\label{a1}

In this appendix we spell out conventions on the fiducial
structures that are used in the main body of the paper and provide
an explicit calculation of holonomy used in the definition of the
Hamiltonian constraint. The discussion on conventions is somewhat
detailed because there has been some confusion due to unfortunate
typos in some of the standard literature on homogeneous
cosmologies.

\subsection{Invariant frames}
\label{a1.1}

In the $k$=1 case the underlying, spatial 3-manifold $M$ is a
3-sphere $\S^3$. It is often convenient to identify it with the
symmetry group $\SU(2)$ which acts on it simply and transitively.
In what follows we will generally do so. Let us denote the
symmetry vector fields on $M$ by $\xi^a_i$. All homogeneous
isotropic tensor fields on $M$ are invariant under diffeomorphisms
generated by $\xi^a_i$. In particular, these vector fields are the
Killing vectors of all 3-metrics considered in this paper.

Let us fix a basis $\tau^i$ in the Lie algebra $\su(2)$,
satisfying $\tau^i \tau^j = \fs{1}{2} \ep^{ij}{}_k \tau^k -
\fs{1}{4} \delta^{ij} \mathbb{I}$ and denote by $k_{ij}$ the
metric on $\su(2)$ for which these $\tau^i$ constitute an
orthonormal basis. In what follows the `internal' or Lie-algebra
indices will be lowered and raised using $k_{ij}$ and its inverse
$k^{ij}$. It follows in particular that $\ep_{ijk} := \ep^{mn}{}_k
q_{mi}q_{jn}$ is a 3-form on $\su(2)$ satisfying $\ep_{ijk}
\ep^{ijk} =6$.

Recall that $SU(2)$ admits a natural, left invariant, Lie-algebra
valued, Cartan 1-form $\omega = g^{-1} dg$. It naturally defines a
co-frame $\w_a^i$ on $M$ via:
\be g^{-1}dg =: \w = \w^i \tau_i \ee
We will denote the dual frame by $\e^a_i$; thus $\e^a_i\, \w_a^j =
\delta_i^j$ and $\e^a_i\, \w_b^i = \delta_b^a$. From the
definition of the natural left-invariant 1-form $\omega$ it
follows that the co-frames $\w_a^i$ and the frames $\e^a_i$
satisfy the relations:
\be \dd\w^i + \fs{1}{2} \ep^i{}_{jk}\, \w^j\wedge\w^k \quad{\rm
and} \quad [\e_i,\,\e_j] = \ep_{ij}{}^k\, \e_k \ee
\emph{These will be our fiducial co-frames and frames on $M$.} The
1-forms $\w_a^i$ and the vector fields $\e^a_i$ are left
invariant. Thus $\xi^a_i$, the infinitesimal generators of left
translations, Lie drag these fields:
\be {\cal L}_{\xi_i}\, \w^j =0, \quad \quad\quad {\cal
L}_{\xi_i}\, \e_j =0\ee
and satisfy the $\su(2)$ commutation relations $[\xi_i, \xi_j] =
\ep_{ij}{}^k \xi_k $. The metric
\be \q_{ab} := \w_a^i\,\w_b^j\, k_{ij}\ee
on $M$ will serve as the fiducial metric on $M$. By inspection
$\xi^a_i$ are Killing fields of $\q_{ab}$. Since they act simply
and transitively on $M$, it follows that $\q_{ab}$ is of constant
curvature. \emph{However, in contrast to one's initial
expectations, it is the metric on a 3-sphere of radius $a=2$
(rather than $a=1$).}%
\footnote{This normalization is fixed by our choice that the
natural left invariant 1-forms $\w_a^i$ be orthonormal. In the
frame formalism it would be awkward and geometrically unnatural to
work with rescaled $\w_a^i$. If one is interested only in metrics
and not frames, on the other hand, one can just as easily work
with an unit 3-sphere. This is the usual choice in
geometrodynamics. In the literature on cosmology the fiducial
metric is sometimes chosen to have unit \emph{scalar curvature}
${}^o\!R$ rather than unit radius; then the 3-sphere radius is $a=
1/\sqrt{6}$.}

The volume of $M$ with respect to $\q_{ab}$ is $V_o= 2\pi^2a^3 =
16\pi^2$ and its scalar curvature is ${}^o\!R= 6/a^2 = 3/2$.
\emph{It will be convenient for us to use the symbol $V_o$}
(rather than the numerical value $16\pi^2$) \emph{to denote the
fiducial volume of $M$. We will also set $\lo = V_o^{1/3}$.} In
the standard charts used in textbooks, the components of $\q_{ab}$
can be expressed as:
\ba ds_o^2 &=&  a^2 \left[\dd\chi^2 + \sin^2\chi\, (\dd \theta^2 +
\sin^2\theta \,\dd \varphi^2)\right]
\nonumber\\
&=& a^2 \left[ \f{\dd r^2}{1-r^2} + r^2 (\dd \theta^2 +
\sin^2\theta \,\dd \varphi^2)\right] \ea
with $a=2$, where $\chi, \theta \in (0, \pi)$,\, $\phi \in
(0,2\pi)$ and $r = \sin \chi\in (0, 1)$.

Finally, the fact that the choice $a=2$ in the fiducial metric is
geometrically natural can also be seen directly in terms of
3-sphere geometries without reference to $\SU(2)$ and the natural
Cartan form thereon. Consider the Euclidean metric and 3-spheres
$\S_{(a)}$ of radius $a$ on $\R^4$. The natural action of the
rotation group ${\rm SO(4)}$ on $\R^4$ leaves each $\S_{(a)}$
invariant. It is the isometry group of the intrinsic metric on
$\S_{(a)}$. These six Killing fields can be naturally divided into
two ${\rm SO(3)}$ sub-Lie-algebras (resulting from self-dual and
anti-self-dual 2-forms on $\R^4$). These are the right and left
rotations and all three right rotations commute with all three
left. In the natural chart $x,y,z,w$ on $\R^4$, these six Killing
fields $K^\pm_i$ can be expressed as:
\ba K^\pm_1 &=& \fs{1}{2}\left( x \partial_y - y \partial_x \pm
z \partial w \mp w \partial z \right)\nonumber\\
K^\pm_2 &=& \fs{1}{2}\left( w \partial_x - x \partial_w \pm
z \partial y \mp y \partial z \right)\nonumber\\
K^\pm_3 &=& \fs{1}{2}\left( y \partial_w - w \partial_y \pm z
\partial x \mp x \partial z \right)\ea
so that $[K^\pm_i, K^\pm_j] = \ep_{ij}{}^k K^\pm_k$ and $[K^+_i,
K^-_j] =0$. Note that because of the first set of commutation
relations, we do not have the freedom to rescale the $K^\pm_i$ by
a constant. The three vectors $K^\pm_i$ in each set are mutually
orthogonal. Now we can ask: On which $\S^3$ is this basis
orthonormal? The answer is: the 3-sphere with radius $a=2$. On
$M$, these are the basis $\{\e^a_i\}$ and $\{ \xi^a_i \}$.

\subsection{An explicit chart}
\label{a1.2}

In the textbook treatments of $\SU(2)$, its elements are often
written as $2\times 2$ matrices:
\ba  g \,=\, \begin{pmatrix}a & -b\\
              b^\star& a^\star
            \end{pmatrix}
          \quad\quad{\rm where}\quad a&=& e^{\fs{i}{2}
(\alpha+\gamma)}\, \cos (\beta/2)\nonumber\\
b&=& e^{-\fs{i}{2} (\alpha-\gamma)}\, \sin (\beta/2) \ea
where $0\le \alpha <2\pi,\, 0\le \beta \le \pi$ and $0\le \gamma <
4\pi$. (The ranges of $\alpha$ and $\gamma$ can be interchanged.)
As is usual with angular coordinates, this chart breaks down at
the poles $\alpha =0, \beta= 0, \gamma =0$ (which corresponds to
the identity, $\mathbb{I}$) and $\alpha=0, \beta= \pi, \gamma=0$
(which corresponds to $-\mathbb{I}$). Nonetheless, as with the
standard spherical polar coordinates, this chart is convenient for
explicit calculations.

In this chart, the left invariant co-frame $\w_a^i$ can be
expressed as:
\ba \w^1 &=& -\cos\g \dd \b - \sin \g \sin\b \dd \a\nonumber\\
\w^2 &=& \sin\g \dd \b - \cos\g \sin\b \dd \a\nonumber\\
\w^3 &=& \dd \g + \cos\beta \dd \a\,  \ea
and the left invariant frame $e^a_i$ as
\ba \e_1 &=& -\cos\g \f{\p}{\p\b} - \f{\sin\g}{\sin
\b}\f{\p}{\p\a} +\f{\cos\b\sin\g}{\sin\b}\, \f{\p}{\p\g}
\nonumber\\
\e_2 &=& \sin\g \f{\p}{\p\b} - \f{\cos\g}{\cos\b}\f{\p}{\p\a} +
\f{\cos\b \cos\g}{\sin\b} \f{\p}{\p\g}\nonumber\\
\e_3 &=& \f{\p}{\p\g}.\ea
Finally the right invariant vector fields ---which Lie drag our
co-frame and the frame-- are given by:
\ba \xi_1 &=& \cos\a \f{\p}{\p\b} - \f{\cos\b\cos\a}{\sin
\b}\f{\p}{\p\a} +\f{\sin\a}{\sin\b}\, \f{\p}{\p\g}
\nonumber\\
&=& -[\cos \a \cos\g -\cos\b\sin\a\sin\g]\e_1 + [\cos\a\sin\g
+\cos\b\sin\a\cos\g]\e_2 + [\sin\a\sin\b] \e_3\nonumber\\
\xi_2 &=& \sin\a \f{\p}{\p\b} + \f{\cos\b\cos\a}{\sin\b}
\f{\p}{\p\a} - \f{\cos\a }{\sin\b} \f{\p}{\p\g}\nonumber\\
&=& -[\cos \g \sin\a +\cos\b\cos\a\sin\g]\e_1 + [\sin\a\sin\g
-\cos\b\cos\a\cos\g]\e_2 - [\cos\a\sin\b] \e_3\nonumber\\
\xi_3 &=& - \f{\p}{\p\a}\nonumber\\
&=& \sin\b\sin\g\, \e_1 + \sin\b\cos\g\, \e_2 -\cos\b \e_3. \ea
Note that although the components of $e^a_i$ and $\xi^a_i$ diverge
on the 2-torus $\beta=0$, this is just an artifact of the failure
of the chart there. All six vector fields are globally
well-defined. In particular the expressions of $\xi^a_i$ as linear
combinations of $e^a_i$ are globally well-behaved.

From the expression of the 1-forms $\w_a^i$ we can readily compute
the components of the fiducial metric:
\be dS_o^2 = \dd \a^2 + \dd \b^2 + \dd \g^2 + 2\dd\a \dd\g \ee
The integral curves of the vector fields $\e^a_3$ (as well as
$\xi^a_3$) are circles which provide a Hopf fibration of $\S^3$.
The quotient $\h{M}$ is topologically $\S^2$ with the induced
metric
\be d\h{s}^2 = \dd\b^2 + \sin^2 \b \dd\a^2 \ee
Thus $\b, \a$ are the standard polar coordinates on the quotient
$\h{M}$ and the induced metric on it is that of an \emph{unit}
2-sphere.

\subsection{Holonomy}
\label{a1.3}

The Hamiltonian constraint involves the field strength $F_{ab}^k$
of the $\SU(2)$-connection
\be A_a^i := c \w_a^i \ee
In LQG, there is no operator corresponding to the connection
itself; only the holonomy operators are well-defined. Therefore,
we need to express the field strength as an appropriate limit of a
suitable holonomy \cite{abl,aps2,aps3}. Now in  $k$=0 cosmologies,
the right and left invariant vector fields on $M$ coincide. Since
they generate translations along the $x,y,z$ directions, they
commute. Hence to calculate the component $F_{32}^k:= \e^a_3\,
\e^b_2\, F_{ab}^k$ of the field strength, we can construct a
square $\Box$ by moving along $z$ and $y$ directions a distance
$\bar\mu$ (as measured by the fiducial metric $\q_{ab}$), evaluate
the holonomy along this loop, divide it by the area enclosed by
the loop and take the limit as the loop shrinks. Because of
homogeneity, $F_{32}^k$ can be evaluated at any point of $M$ and
because of isotropy, $F_{32}^k$ determines $F_{ab}^k$ completely.

In the present $k$=1 case we wish to follow a similar procedure.
Again, $F_{32}^k$ can be evaluated at any point of $M$ and fully
determines $F_{ab}^k$. However, now the vector fields $\e^a_3$ and
$\e^a_2$ do not commute. Therefore, as explained in the main text,
we can not form the desired loop $\Box_{ij}$ by moving along 4
segments of their integral curves. However, since the left
invariant vector fields $\e^a_i$ do commute with the right
invariant vector fields $\xi^a_i$, we \emph{can} construct the
desired closed loop $\Box_{ij}$ using integral curves of $\e^a_i$
and $\xi^a_j$.

For definiteness, let us start with the point $(\a =0, \b=\pi/2,
\g= 0)$. At this point, $\xi_3^a = -\e_2^a$ and we can calculate
the component $\e^a_3\,\,\e^b_2\, F_{ab}$ using the loop
$\Box_{32}$ constructed from the composition of the following four
segments:
\begin{quote}
Segment 1: move from $(0,\pi/2,0)$ to $(0,\pi/2,\lambda\lo)$ along
the integral curve of $\e_3 = \p/\p\g$;\\
Segment 2: move along $(0,\pi/2,\lambda\lo)$ to
$(\lambda\lo,\pi/2,\lambda\lo)$ along
the integral curve of $-\xi_3^a = \p/\p\a$; \\
Segment 3: move from $(\lambda\lo,\pi/2,\lambda\lo)$ to
$(\lambda\lo,\pi/2,0)$ along the integral curve of $-\e_3 = -\p/\p\g$;\\
Segment 4: return from $(\lambda\lo,\pi/2,0)$ to the point of
departure $(0,\pi/2,0)$ along the integral curve of $\xi_3^a =
-\p/\p\a$.
\end{quote}

It is straightforward to calculate the holonomies along these four
segments. One obtains
\be h_1 = e^{\lambda c\tau_3}, \quad h_2 = e^{-\lambda
c(\sin\lambda\,\tau_1 + \cos\lambda \tau_2)},\quad h_3 =
e^{-\lambda c \tau_3}\quad h_4 = e^{\lambda \tau_2} \ee
The holonomy $h_{\Box_{32}}$ around the loop is just the
composition $h_4 h_3 h_2 h_1$. A simple calculation yields:
\be \e^a_3\,\, \e^b_2\, F_{ab}^k = \lim_{\lambda\rightarrow 0}
\f{4}{\lambda^2\lo^2}\, \tr(h_{{\Box}_{32}}\tau^k) =
-\f{1}{\lo^2}\,(c^2 - c\lo)\delta^k_1\ee
Finally, using isotropy and homogeneity, one concludes:
\ba F_{ab}^k &=& - \lim_{\lambda\rightarrow 0}\,\,
\f{1}{\lambda^2\lo^2}\,
(\tr\, h_{\Box_{ij}}\tau^k)\,\w_a^i\,\,\w_b^j\nonumber\\
&=& - \lim_{\lambda\rightarrow 0}\,\,
\f{1}{\lambda^2\lo^2}\,[\sin^2\,\lambda(c-\lo/2) - \sin^2
(\lambda\lo/2) ]\,  \ep_{ij}{}^k \,\w_a^i\,\,\w_b^j\nonumber\\
&=& \lo^{-2}\,\,(c^2 - c\lo) \, \ep_{ij}{}^k\, \w_a^i\,\,\w_b^j
\ea

The expression of the holonomy in the second step is used in the
discussion of the Hamiltonian constraint in section \ref{s2} of
the main text.

\end{appendix}

\end{document}